\documentclass[aps,pra,twocolumn,amssymb,amsfonts,floatfix]{revtex4}
\usepackage{bm}
\usepackage{dcolumn}
\usepackage{bm}
\usepackage{mathbbol}
\usepackage{graphicx}
\usepackage{epstopdf} 
\usepackage{amsmath,amsthm}
\usepackage{times}
\usepackage{color}
\usepackage{lipsum}

\begin{document}
\title{Power-law decay exponents: a dynamical criterion for predicting thermalization} 

\author{Marco T\'avora, E. J. Torres-Herrera, and Lea F. Santos}
\affiliation{Department of Physics, Yeshiva University, New York, New York 10016, USA}
\affiliation{Instituto de F{\'i}sica, Universidad Aut\'onoma de Puebla, Apt. Postal J-48, Puebla, Puebla, 72570, Mexico}

\date{\today}

\begin{abstract}
From the analysis of the relaxation process of isolated lattice many-body quantum systems quenched far from equilibrium, we deduce a criterion for predicting when they are certain to thermalize. It is based on the algebraic behavior $\propto t^{-\gamma}$ of the survival probability at long times. We show that the value of the power-law exponent $\gamma$ depends on the shape and filling of the weighted energy distribution of the initial state. Two scenarios are explored in details: $\gamma \ge 2$ and $\gamma <1$.  Exponents $\gamma \ge 2$ imply that the energy distribution of the initial state is ergodically filled and the eigenstates are uncorrelated, so thermalization is guaranteed to happen. In this case, the power-law behavior  is caused by bounds in the energy spectrum. Decays with $\gamma < 1$ emerge when the energy eigenstates are correlated and signal lack of ergodicity. They are typical of systems undergoing localization due to strong onsite disorder and are found also in clean integrable systems. 
\end{abstract}


\maketitle


\section{Introduction}
\label{sec:introduction}

Equilibrium quantum physics can be effectively described with the framework of quantum statistical mechanics, but the dynamics that lead to equilibration  is far less understood. Recently, the analysis of nonequilibrium quantum dynamics has been stimulated by the enormous progress in experimental techniques, particularly the manipulation of ultracold atomic gases~\cite{Trotzky2012,Fukuhara2013}, trapped ions~\cite{Jurcevic2014,Richerme2014}, and nuclear magnetic resonance (NMR) platforms~\cite{Cappellaro2007,Kaur2013}, where coherent evolutions can be studied for long times. Questions that have been at the forefront of these investigations include  the characterization of the dynamics of isolated lattice many-body quantum systems at different time scales and whether they can or cannot eventually thermalize.

The onset of thermalization in isolated quantum systems is intimately attached to the onset of quantum chaos, which causes the uniformization of the eigenstates~\cite{ZelevinskyRep1996,Santos2010PRE,RigolSantos2010,Santos2010PREb,Santos2011PRL,Santos2012PRL,Santos2012PRE,Borgonovi2016,Torres2016BJP} and guarantees the coincidence of infinite-time averages and thermodynamic averages of few-body observables~\cite{Deutsch1991,Srednicki1994,Rigol2008,rigol09STATa,rigol09STATb,Torres2013,He2013,Torres2014PRE,AlessioARXIV}. In real systems, however, where only few-body interactions exist, even in the presence of level repulsion, the eigenstates are not truly chaotic (pseudo-random vectors), as in full random matrices. Nearly random vectors tend to emerge only away from the edges of the spectrum.  

One of the main approaches of the studies of thermalization in realistic finite systems is the use of scaling analysis to identify for which systems and in which regions of the spectrum, chaotic eigenstates emerge and statistical descriptions become valid. However,  the range of system sizes that can be reached numerically is limited, which prevents effective scaling analysis. An alternative is to directly access the thermodynamic limit using linked-cluster computational methods for some initial states~\cite{Rigol2014,Rigol2016}. Another option is to look for dynamical properties from which one can infer the structure of the initial state and use it to determine whether the system will or not thermalize. This is the approach that we introduced in Ref.~\cite{TavoraARXIV} and further extend here.

The onset of thermalization can be investigated by studying the decay at long times of the survival probability. It corresponds to the probability for finding the system still in its initial state at a later time $t$. At long times, no matter how fast the decay may initially be, the survival probability necessarily exhibits a power-law behavior $\propto t^{-\gamma}$. The value of $\gamma$ varies according to the system and initial state, but the unavoidable onset of the algebraic decay is independent of whether the finite system is integrable or chaotic, disordered or not, interacting or noninteracting. 

The exponent $\gamma$ depends on the shape and filling of the weighted energy distribution of the initial state, which is here referred to as local density of states (LDOS). The ergodic filling of the LDOS assures that the initial state is highly delocalized and similar to a chaotic state, which is a sufficient condition for thermalization~\cite{Santos2011PRL,Santos2012PRL,Santos2012PRE,Torres2013,He2013,Rigol2016}. Therefore, the value of $\gamma$ can be used as a criterion for identifying which systems and for which initial states thermalization is guaranteed to take place. 

We focus on two classes of exponents:
\begin{itemize}
  \item {\bf Case 1} corresponds to $\gamma \ge 2$. It is related with the presence of bounds in the energy spectrum~\cite{Khalfin1958,Nussenzweig1961,Ersak1969,Fleming1973,Knight1977,Fonda1978,Sluis1991,Campo2011,Campo2016,MugaBook,Peshkin2014}. This cause for the algebraic decay has been much explored in the context of continuous models. Here, we show that energy bounds are also the main cause of the power-law decay in lattice many-body quantum systems when the initial state has an ergodically filled LDOS.  In realistic lattice models with two-body interactions, $\gamma=2$, while in non-physical systems with the simultaneous interactions of many particles, the exponent can reach the limit of $\gamma= 3$ \cite{TavoraARXIV,Torres2016}. Exponents in this range of values anticipate thermalization.
  \item {\bf Case 2} refers to $0<\gamma < 1$. It occurs when the LDOS is sparse, which indicates lack of ergodicity. The decay exponent is related to the presence of correlations in the eigenstates of the Hamiltonian~\cite{Chalker1988,Chalker1990,Ketzmerick1992,Huckestein1994,Huckestein1999,Cuevas2007,Kravtsov2011,Evers2008,DeLuca2013,Torres2015,Torres2015BJP}. It  has been studied in the context of Anderson localization~\cite{Ketzmerick1992,Huckestein1994,Huckestein1999,Cuevas2007,Kravtsov2011} and more recently in interacting systems with onsite disorder~\cite{Torres2015,Torres2015BJP,TorresARXIV}. Here, we show that exponents $\gamma<1$ emerge also in noninteracting integrable models without disorder. The methods that have been developed to extract the value of $\gamma$ in disordered systems apply also for these clean models. 
\end{itemize}

There are integrable systems studied in the literature \cite{Venuti2010,Venuti2011,Happola2012} for which we find $1\le \gamma < 2$. This range of exponents is obtained also in disordered interacting systems in the chaotic domain, although not at the point of maximum delocalization of the eigenstates~\cite{TavoraARXIV,TorresARXIV}. The origin of the algebraic decay for these values of $\gamma$ is not yet clear. 

We note that the emergence of power-law decays has been observed also for different physical observables, especially in works about disordered systems. In most of these cases, the values of the exponents have not been analytically justified, as we do here. A  discussion about the power-law route to thermal equilibrium is found, for instance, in \cite{Khatami2012}.

This paper is organized as follows. Section II introduces the survival probability and describes its behavior at short and intermediate times. Section III summarizes the theory associated with the behavior of the survival probability at long times. Section IV illustrates these behaviors for a spin-1/2 system under different choices of parameters and initial states.  Final remarks are given in Sec.~V. Details about the calculations are found in the Appendixes A, B, and C.


\section{Survival Probability and LDOS}

The time evolution of an initial nonstationary state may be viewed as follows. Suppose that the system is prepared at $t=0$ in some initial state $|\Psi (0)\rangle$, which is an eigenstate of a Hamiltonian $H_0$. The dynamics is triggered by rapidly changing (quenching) the Hamiltonian to a new final Hamiltonian $H$,
\begin{equation}
H_0 \longrightarrow H=H_0+gV,
\label{eq:quench}
\end{equation}
where $g$ is the perturbation strength. The probability for finding the system at time $t$ still in state $|\Psi (0)\rangle$ is known as the survival probability and is given by
\begin{equation}
F(t)=|A(t)|^2 \equiv |\langle \Psi (0)| e^{ - iHt} |\Psi (0)\rangle |^2 ,
\label{eq:F}
\end{equation}
 where $A(t) $ is the survival amplitude. $F(t)$ is also referred to as nondecay probability, return probability, or fidelity between the initial state and the evolved one. The term Loschmidt echo is not appropriate in this case, since no time reversal (``echo'') is  involved. 

By projecting the initial state on the eigenstates $|\psi_{\alpha}\rangle$ of $H$ and substituting it into Eq.~(\ref{eq:F}), we obtain
\begin{eqnarray}
F(t) = \left| \sum\limits_\alpha  | C_\alpha ^{(0)} |^2 e^{ - i E_\alpha t } \right|^2 =  \left| \int dE\, e^{ - iEt} \rho_0(E) \right|^2,
\label{saitorho}
\end{eqnarray} 
where $C_{\alpha}^{(0)}=\langle \psi_{\alpha} | \Psi(0) \rangle$ are the overlaps and
\begin{equation}
\rho_0(E)  \equiv \sum\limits_\alpha  | C_\alpha ^{(0)}|^2 \delta (E - E_\alpha )
\end{equation} 
is the LDOS (also known in nuclear physics as strength function). The survival amplitude is the Fourier transform of the LDOS, or equivalently, $A(t)$ is the characteristic function of the weighted energy distribution. All information about the evolution of $F(t)$ is contained in $\rho_0(E)$.

The energy and variance of the initial state are important elements in the description of the dynamics. They are respectively given by
\begin{equation}
E_0= \langle \Psi (0)| H |\Psi (0)\rangle = \sum_{\alpha} |C_{\alpha}^{(0)}|^2 E_{\alpha},
\label{energyinitialstate}
\end{equation}
and 
\begin{equation}
\sigma_0^2=\sum_{\alpha} |C_{\alpha}^{(0)} |^2 (E_{\alpha} - E_0)^2.
\label{stinitialstate}
\end{equation}

The decay of $F(t)$ shows different behaviors at different time scales. 
For very short times, $t \ll \sigma _0^{ - 1}$, the decay is quadratic, as observed experimentally~\cite{Wilkinson1997}. After this universal quadratic behavior, the decay depends on the nature and strength of the perturbation. In lattice many-body quantum systems with two-body interactions and a unimodal LDOS, if the perturbation is strong, the decay can be exponential or even Gaussian~\cite{Torres2014PRE,Torres2014PRA, Torres2014NJP,Torres2014PRAb,TorresKollmar2015,TorresProceed}. This second behavior holds for $\sigma _0^{ - 1} \lesssim t \lesssim {t_P}$, where $t_P$ corresponds to the moment of the onset of the power-law decay. At long times, $t \gtrsim {t_P}$, the dynamics is necessarily algebraic, $F(t) \propto t^{-\gamma}$. This work is mainly concerned with this last time regime.


\subsection{Short and intermediate time scales: $\pmb{t < t_P}$}
\label{Sec:short}

By Taylor expanding the phase factor in Eq.~(\ref{saitorho}), it is straightforward to show that the survival probability at very short times, $t \ll \sigma _0^{ - 1}$, is quadratic in $t$,
\begin{eqnarray}
F(t) &&\approx \left| e^{-i E_0 t}\left[ \sum_\alpha  | C_\alpha^{(0)} |^2 - i \sum_\alpha  | C_\alpha ^{(0)}|^2(E_\alpha  - E_0)t \right. \right.\nonumber\\
&&\left. \left.  - \frac{1}{2} \sum_\alpha  | C_\alpha ^{(0)}|^2 (E_\alpha  - E_0)^2 t^2 \right] \right|^2  \nonumber\\
&& \approx  1 - \sigma _0^2 t^2 ,
\label{quadractic}
\end{eqnarray}
independently of the initial state and the Hamiltonian $H$.

For intermediate times, $\sigma _0^{ - 1} \lesssim t \lesssim {t_P}$, the behavior of $F(t)$ depends on the shape of the LDOS, which, in turn, depends on the strength of the perturbation. In systems with two-body interactions, the density of states is Gaussian~\cite{Brody1981,Kota2001,Santos2012PRE,Zangara2013}. In this scenario, the LDOS, which is a delta function for $g=0$, broadens as the strength of the perturbation increases. When the perturbation $gV$ is stronger than the mean level spacing (Fermi golden rule regime), the LDOS becomes a Lorentzian (also known as Breit-Wigner) of width $\Gamma _0$,
\begin{equation}
\rho _0(E) = \frac{1}{2\pi } \frac{\Gamma _0}{\left( E_0 - E \right)^2 + \Gamma _0^2/4} .
\end{equation}
The Fourier transform of the Lorentzian leads to the exponential behavior
\begin{equation}
F(t) = \exp ( - \Gamma _0 t).
\end{equation}
As the perturbation further increases, the LDOS stretches and eventually reaches a Gaussian shape (different functions are used to fit the intermediate regime between the Lorentzian and Gaussian form~\cite{Frazier1996,Flambaum2001a,Chavda2004,KotaBook,Torres2014PRAb}). The Gaussian LDOS that emerges when $g\to 1$, 
\begin{equation}
\rho _0(E) = \frac{1}{\sqrt {2\pi \sigma _0^2} } \exp \left[ - \frac{(E - E_0)^2}{2\sigma _0^2} \right] ,
\label{eq:GaussLDOS}
\end{equation}
reflects the density of states, which, as said above, is also Gaussian. This is the maximum spreading of the initial state. In this case, the survival probability decay is Gaussian,
\begin{equation}
F(t) = \exp ( - \sigma _0^2 t^2).
\label{fidgaussianinterm}
\end{equation}
Notice that whether the decay is exponential or Gaussian depends on the strength of the perturbation and not on the regime, integrable or chaotic, of the final Hamiltonian. Gaussian and Lorentzian LDOS can be found in quenches to both chaotic and also integrable Hamiltonians~\cite{Santos2012PRL,Santos2012PRE,Borgonovi2016,Torres2014PRA, Torres2014NJP,Torres2014PRAb,TorresKollmar2015,TorresProceed}.

There are special situations where the decays can be even faster than Gaussian. This happens, for instance, when the LDOS is bimodal and the decay is dictated by the distance between the peaks~\cite{Torres2014PRAb}. Another example corresponds to systems with random many-body interactions, the extreme case being that of full random matrices, where the density of states and also the LDOS have a semicircular shape~\cite{Wigner1955,Torres2014PRA,Torres2014NJP,Torres2014PRAb},
\begin{equation}
\rho_0(E) = \frac{1}{2\pi \sigma _0^2} \sqrt{  (2 \sigma_0)^2 -E^2  } .
\label{eq:semicircle}
\end{equation}
Full random matrices are matrices filled with random numbers. Their only constraint is to satisfy the symmetries of the system they try to represent~\cite{Guhr1998}. They are unphysical, because they imply that all the particles interact simultaneously. However, they are useful to establish bounds for the speed of the evolution. The Fourier transform of the semicircle gives the following analytical expression for the survival probability~\cite{Torres2014PRA, Torres2014NJP,Torres2016}
\begin{equation}
F(t)= \frac{[{\cal J}_1(2\sigma_0 t)]^2}{\sigma_0^2 t^2} ,
\label{eq:FRMdecay}
\end{equation}
where ${\cal J}_1$ is the Bessel function of the first kind. Equation~(\ref{eq:FRMdecay}) gives the fastest possible decay of the survival probability for lattice many-body quantum systems with a unimodal LDOS.


\section{Long-time scales: $\pmb{t > t_P}$}
\label{LT}

While for $t<t_P$, the dynamics can be very fast depending on the envelope of the LDOS, at long times the decay of the survival probability slows down and necessarily shows a power-law behavior,
\begin{equation}
F(t) \propto t^{ - \gamma } \hspace{0.7 cm}  (\gamma  > 0).
\label{generalFlambdalt2}
\end{equation}
The theoretical causes for the algebraic decay corresponding to Case 1 ($\gamma  \geq 2$) and Case 2 ($\gamma  < 1$) are explained below. Numerical examples, as well as a brief discussion about the intermediate region $1 \le \gamma  < 2$, are given in Sec.~\ref{Sec:chaos}.


\subsection{Case 1: $\pmb{\gamma \ge 2}$ (Ergodically Filled LDOS)}
\label{subcase1}

Any real quantum system necessarily has a lower bound in the energy spectrum, which we denote by $E_{low}$. Taking this bound into account in the LDOS, that is ${\rho _0}(E<E_{low})=0$, and using the Paley-Wiener theorem, Khalfin showed in 1958~\cite{Khalfin1958} that 
the survival probability at long times has to decay more slowly than exponentially~\cite{Khalfin1958,Fonda1978}. The behavior should become $F(t) \propto \exp ( - c t^q )$, with $c>0$ and $q<1$.  This study was done for LDOS that were absolutely integrable functions, that is~\cite{Fock1947}
\begin{equation}
\int_{\Delta E} \rho_0(E')dE'\mathop  \to \limits^{\Delta E \to 0} 0  ,
\label{eq:Lebesgue}
\end{equation}
where $\Delta E$ is any interval inside the spectrum.

Asymptotic analyses have actually shown that the decay of $F(t)$ becomes power-law at long times and that the exact value of the exponent $\gamma$ in Eq.~(\ref{generalFlambdalt2}) depends on how the LDOS decays to zero at the bounds of the spectrum~\cite{Erdelyi1956,Urbanowski2009}. 
Assuming that $\rho_0(E)$ is absolutely integrable and that its derivatives exist and are continuous in $[E_{low}, \infty]$, two cases are singled out:

(i) If the LDOS is such that
\[
\mathop {\lim }\limits_{E \to E_{low}} \rho_0 (E) > 0 ,
\] 
the survival probability decays as 
\begin{equation}
F(t) \propto t^{-2}.
\end{equation}
Gaussian and Lorentzian LDOS belong to this class. The Gaussian LDOS with exponential tails considered in nuclear shell models~\cite{Frazier1996} also fall in this category. Details on how to obtain the $t^{-2}$ decay are shown in Appendix~\ref{RigRes}. There, we consider the general case, where both bounds are present, the lower, $E_{low}$, and the upper one, $E_{up}$.

(ii) If the LDOS goes to zero at $E_{low}$, that is
\begin{equation}
\rho_0(E) = (E-E_{low})^{\xi} \eta(E) ,
\end{equation}
with
\[
\lim _{E\rightarrow E_{low}} \eta(E)>0 , 
\]
and $0<\xi<1$, and if the  derivatives of $\eta(E)$ exist and are continuous in $[E_{low}, \infty]$, then the decay is given by
\begin{equation}
F(t) \propto t^{-2(\xi+1)}.
\label{eq:xi_gamma}
\end{equation}
Hence, apart from how the LDOS approaches the energy bound, its exact shape does not play an important role in the long-time decay of $F(t)$.

Examples of Case 1 (ii) for continuous models describing a trapped particle in a inverse-square potentials are found in~\cite{Martorell2008,Torrontegui2010}. 
The semicircle LDOS shown in Eq.~(\ref{eq:semicircle}) also belongs to Case 1 (ii). For it, one has $\xi =1/2$, $E_{low}= - 2\sigma_0$, and 
\[
\eta(E) = \frac{\sqrt{ 2\sigma_0 - E }}{2 \pi \sigma_0^2},
\]
which leads to
\begin{equation}
F(t) \propto t^{-3}.
\label{eq:Fsemi}
\end{equation}
This result can also be derived directly from the analytical expression of the survival probability given in Eq.~(\ref{eq:FRMdecay}). For $t\gg \sigma_0^{-1}$, one finds that
\begin{equation}
F(t \gg \sigma _0^{ - 1}) \to \frac{1 - \sin (4 \sigma_0t)}{2\pi \sigma_0^3 t^3}.
\label{eq:FRMasymptotic}
\end{equation}
The value $\gamma = 3$ should therefore be the upper bound for the power-law exponent of $F(t)$ of finite lattice many-body quantum systems.

\subsubsection{Thermalization and  $\gamma \geq 2$}
The results described above are valid for continuous functions. In finite lattice many-body quantum systems, where the spectrum is discrete, we expect the power-law exponent to approach values $\gamma \geq 2$ when the LDOS is ergodically filled. By this we mean that the initial state, projected on the energy eigenbasis, is very similar to a pseudo-random vector, so it samples most of the energy eigenbasis with energy within $\sigma_0$ (most ${C_\alpha ^{(0)}}$ are nonzero) without any preference (${C_\alpha ^{(0)}}$ are close to uncorrelated random numbers). As a result of the ergodicity, the LDOS is well approximated by an absolutely integrable function.

Ergodicity is certainly satisfied for arbitrary initial states projected onto the eigenstates of full random matrices. Since all of these eigenstates are chaotic (pseudo-random) vectors, so is the projected initial state. Full random matrices, however, do not describe realistic systems. For the latter, where only few-body interactions exist, the Hamiltonian matrices are sparse, random elements may not even be present, and the density of states is Gaussian instead of semicircular.  Yet, in the chaotic regime, these systems still follow random matrix statistics, that is, away from the edges of the spectrum, there occurs level repulsion and the eigenstates are very similar to random vectors. In the case of a strong perturbation that quenches the initial Hamiltonian into such final chaotic Hamiltonians, the LDOS of the initial state will also be very well filled, since $|\Psi(0)\rangle$ is projected onto nearly random vectors~\cite{Borgonovi2016}. 

To verify whether the LDOS is ergodically filled, one uses quantities that measure the level of delocalization of the initial state~\cite{Izrailev1990,ZelevinskyRep1996,Gubin2012}. A commonly employed one is the participation ratio, defined as
\begin{equation}
\mbox{PR}_0 \equiv \frac{1}{\sum_{\alpha} |C_{\alpha}^{(0)}|^4}.
\label{eq:PR}
\end{equation}
A large value of $\mbox{PR}_0$ indicates that the initial state is delocalized in the energy eigenbasis $|\psi_{\alpha}\rangle$. For chaotic (pseudo-random) states, $\mbox{PR}_0 \propto {\cal D}$, where ${\cal D}$ is the dimension of the Hamiltonian matrix.

The value of the PR can be calculated directly from Eq.~(\ref{eq:PR}) by using exact diagonalization or from the survival probability after saturation. Since the treated systems are finite, $F(t)$ eventually saturates to a finite positive value. From Eq.~(\ref{saitorho}), one sees that 
\[
F(t) = \sum_{\alpha} |C_{\alpha}^{(0)} |^4 + \sum_{\alpha \neq \beta} |C_{\alpha}^{(0)} |^2 |C_{\beta}^{(0)} |^2 e^{i (E_{\alpha} - E_{\beta}) t} .
\]
The time average of the second term can be dropped for large times, provided the system does not have an excessive number of degeneracies.  The infinite time average of the survival probability is then
\begin{equation}
\overline{F}=\sum_{\alpha} |C_{\alpha}^{(0)} |^4 \equiv \mbox{IPR}_0,
\label{Eq:saturationF}
\end{equation}
where IPR stands for the inverse of the participation ratio. For finite lattice systems, $\mbox{IPR}_0 \ne 0$.

Quantum chaos and the onset of thermalization are directly linked~\cite{ZelevinskyRep1996,Borgonovi2016}. An ergodically filled LDOS guarantees that the initial state will thermalize~\cite{ZelevinskyRep1996,Borgonovi2016,Santos2011PRL,Torres2013,He2012,Rigol2016}. In this case, the diagonal entropy $S_d$ \cite{Polkovnikov2011}, which is the entropy that characterizes the system after equilibration, and the thermodynamic entropy $S_{th}$ coincide \cite{Santos2011PRL}. 

The diagonal entropy is defined as
\begin{equation}
S_d = - \sum_{\alpha}  |C_{\alpha}^{(0)} |^2 \ln  |C_{\alpha}^{(0)} |^2 .
\label{d-entropy}
\end{equation}
It is the Shannon (information) entropy~\cite{ZelevinskyRep1996} of the initial state written in the energy eigenbasis. As shown in~\cite{Santos2011PRL,Polkovnikov2011}, $S_d$ can be written as the sum of a smooth and a fluctuating part. The smooth part approaches the microcanonical entropy as the system size increases, which in turn coincides with the canonical entropy $S_{can}$ when the system is large. The fluctuating part  becomes negligible for large system sizes when the LDOS is a smooth function of energy, which happens when $\rho_0(E)$ is ergodically filled. The approach of $S_d$ to the thermodynamic entropy as the system size increases was indeed shown numerically in Ref.~\cite{Santos2011PRL} for initial states with energies away from the edges of the spectrum and evolving according to the same chaotic Hamiltonians that are investigated in Sec.~\ref{Sec:gamma2}.

\subsubsection{Time scales}
\label{sec:timescales}

We identify three time scales associated with the distinct behaviors of the survival probability: $t \ll \sigma _0^{ - 1}$,  $\sigma _0^{ - 1} \lesssim t \lesssim {t_P}$, and $t \gtrsim {t_P}$. The time for $F(t)$ to saturate and simply fluctuate around $\text{IPR}_0$ depends on the different behaviors encountered during the evolution. It should have a strong dependence on the width of the LDOS and on the value of $\text{IPR}_0$.

For chaotic initial states, we expect the saturation time to be smaller than the Heisenberg time, $t_{H} \equiv 2\pi/\delta E $, where $\delta E$ is the mean spacing between energy eigenvalues. $t_{H} $ corresponds to the interval after which, due to the energy-time uncertainty principle, the system starts to ``feel'' the discreteness of the spectrum~\cite{Chirikov1985}. This time is large in many-body quantum systems and it grows exponentially with system size.


\subsection{Case 2:  $\pmb{\gamma < 1}$ (Sparse LDOS)}
\label{subcase2}

A sparse LDOS signals the presence of correlated nonchaotic eigenstates in the final Hamiltonian. These states appear in disordered systems that undergo spatial localization due to strong onsite disorder~\cite{Chalker1988,Chalker1990,Ketzmerick1992,Huckestein1994,Huckestein1999,Cuevas2007,Kravtsov2011,Evers2008,DeLuca2013,Torres2015,Torres2015BJP,TorresARXIV}. In this case, the power-law exponent of the survival probability is $\gamma<1$. As we show in Sec.~\ref{XX}, this picture occurs also for noninteracting integrable models without disorder. 

Naturally, the spectrum remains bounded also in these nonchaotic disordered and clean models. However, the exponent of the power-law decay due to correlations is smaller than that caused by the energy bounds, so it is the correlations that determine the behavior of $F(t)$ at long times.

The survival probability can be expressed in terms of the correlation function ${\cal C}(\omega )$ as follows,
\begin{eqnarray}
&&F(t)  = \int_{ - \infty }^\infty  d\omega e^{i\omega t} {\cal C}(\omega ) , \nonumber \\
&&{\cal C}(\omega ) \equiv \sum\limits_{\alpha ,\beta } | C_\beta ^{(0)}|^2 |C_\alpha ^{(0)}|^2 \delta (E_{\alpha } - E_{\beta } - \omega ) .
\label{FtfromComega}
\end{eqnarray}
The long-time behavior of $F(t)$ is dominated by small $\omega$. A power-law decay with $\gamma<1$ emerges at large $t$ when~\cite{Chalker1988,Chalker1990,Ketzmerick1992,Huckestein1994,Huckestein1999,Cuevas2007,Kravtsov2011} 
\begin{equation}
{\cal C}(\omega  \to 0) \propto \omega ^{\gamma  - 1}.
\end{equation}
The value of $\gamma$ indicates the level of correlations between the components $|C_\alpha ^{(0)}|^2$ and thus also between the eigenstates. 

A sparse LDOS is the consequence of a nonergodic initial state, which samples only a portion of the Hilbert space. In this case,
\begin{equation}
\mbox{IPR}_0 \propto {\cal D}^{-D_2}
\end{equation}
with $D_2<1$. The exponent $D_2$ coincides with the power-law exponent of $F(t)$ when $\gamma<1$.  Thus, $\gamma$ can be obtained either from the decay of the survival probability or from the scaling analysis of $\mbox{IPR}_0$,  as extensively done in studies of Anderson localization~\cite{Ketzmerick1992,Huckestein1994,Huckestein1999} and, more recently, many-body localization~\cite{Torres2015,Torres2015BJP,TorresARXIV}.

When the initial state is ergodic, $|C_\alpha ^{(0)}|^2$ are approximately normalized random variables and  $D_2 \rightarrow 1$. Notice that at this point, the power-law decay of $F(t)$ is not determined by correlations anymore, so the scaling analysis of $\mbox{IPR}_0$ can no longer be used to derive the exponent of the algebraic decay.

\section{Results for Spin-1/2 Models}
\label{Sec:chaos}

The general results discussed in the previous section are illustrated here for finite one-dimensional lattice many-body quantum systems described by spin-1/2 models. The Hamiltonian is given by,
\begin{eqnarray}
&&H = H_h +  H_{NN} + \lambda H_{NNN} 
\label{ham}  ,\\
&& H_h =  \sum_{n=1}^L  h_n  S_n^z   \;, \nonumber \\
&& H_{NN} = \sum_{n} J \left(S_n^x S_{n+1}^x + S_n^y S_{n+1}^y +\Delta S_n^z S_{n+1}^z \right) \;,
\nonumber \\
&& H_{NNN} = \sum_{n} J \left(S_n^x S_{n+2}^x + S_n^y S_{n+2}^y +\Delta S_n^z S_{n+2}^z \right) \;.
\nonumber 
\end{eqnarray}
Above, $\hbar=1$, $S^{x,y,z}_n$ are the spin operators on site $n$, and $L$ is the total even number of sites in the chain. The amplitudes $h_n$ are random numbers from a uniform distribution $[-h,h]$, where $h$ is the disorder strength. The system is clean when $h=0$. The Hamiltonian contains nearest-neighbor (NN) and possibly also next-nearest-neighbor (NNN) couplings. The coupling strength $J$, the anisotropy parameter $\Delta$, and the ratio $\lambda$ between NNN and NN couplings are positive. The sums in $ H_{NN} $ and $ H_{NNN} $ go from $n=1$ to $n=L-1$ when the chain has open boundaries and up to $L$ when it has periodic boundaries. The energy scale is set by $J=1$. The total spin in the $z$-direction, ${\cal S}^z$, is conserved. We analyze the largest subspace, where ${\cal S}^z=0$ and the dimension is ${\cal D}=L!/(L/2)!^2$.

Hamiltonian (\ref{ham}) presents the following limits:

(i) It is a noninteracting clean integrable model when $h, \Delta,\lambda=0$. In this case, it is referred to as the $XX$ model. When the couplings in the $x$ and $y$-directions have different strengths, the Hamiltonian represents the $XY$ model.

(ii) It is an interacting clean integrable model, referred to as $XXZ$ model, when $h, \lambda=0$.

(iii) When $\lambda=0$, $\Delta<1$, and $0<h<1$, the spectrum shows level repulsion with the level spacing distribution coinciding with the Wigner-Dyson distribution~\cite{Avishai2002, Santos2004,SantosEscobar2004,Dukesz2009}, as typical of chaotic systems~\cite{Guhr1998,Gubin2012}. In contrast, the levels can cross and many-body localization eventually takes place when the disorder becomes strong \cite{SantosEscobar2004,Dukesz2009,Santos2005loc,Pal2010}.

(iv) When $h=0$, $\Delta<1$, and $\lambda \lesssim1$,  the spectrum again shows level repulsion~\cite{Kudo2005,Santos2009JMP}.

We consider as initial states, site-basis vectors, where the spin on each site points either up or down in the $z$-direction. They include the N\'eel state,
\begin{eqnarray}
&& |\text{NS} \rangle = | \downarrow \uparrow \downarrow \uparrow \downarrow \uparrow \downarrow \uparrow  \ldots \rangle , \nonumber \\
&& E_0 =  \sum_{n }  \frac{(-1)^n h_n}{2} +\frac{J\Delta}{ 4}  [ -(L-1) + (L-2)\lambda ] ,\nonumber \\
&&\sigma_{\text{0}}  = \frac{J}{2} \sqrt{L-1}, 
\label{eq:NS} 
\end{eqnarray}
and the domain wall state,
\begin{eqnarray}
&&|\text{DW} \rangle = |\uparrow \uparrow \uparrow \ldots \downarrow \downarrow  \downarrow  \ldots \rangle ,\nonumber \\
&& E_0 =  \sum_{n }  \frac{(-1)^{\lfloor \frac{2(n-1)}{L} \rfloor } h_n}{2} + \frac{J\Delta}{4} [(L-3) + (L-6)\lambda] ,\nonumber \\
&&\sigma_{\text{0}}  = \frac{J}{2} \sqrt{1+2\lambda^2}  .
\label{eq:DW} 
\end{eqnarray}
These are important states in magnetization. They are often used in theoretical studies of quench dynamics and are accessible to experiments with optical lattices~\cite{Trotzky2008,Schreiber2015}.  They are eigenstates of the initial Hamiltonian $H_0$, where $h, \lambda=0$ and $ \Delta \rightarrow \infty$.

\subsection{Power-law exponent $\pmb{\gamma \ge 2}$}
\label{Sec:gamma2}

\begin{figure}[htb]
\centering
\includegraphics*[width=3.2in]{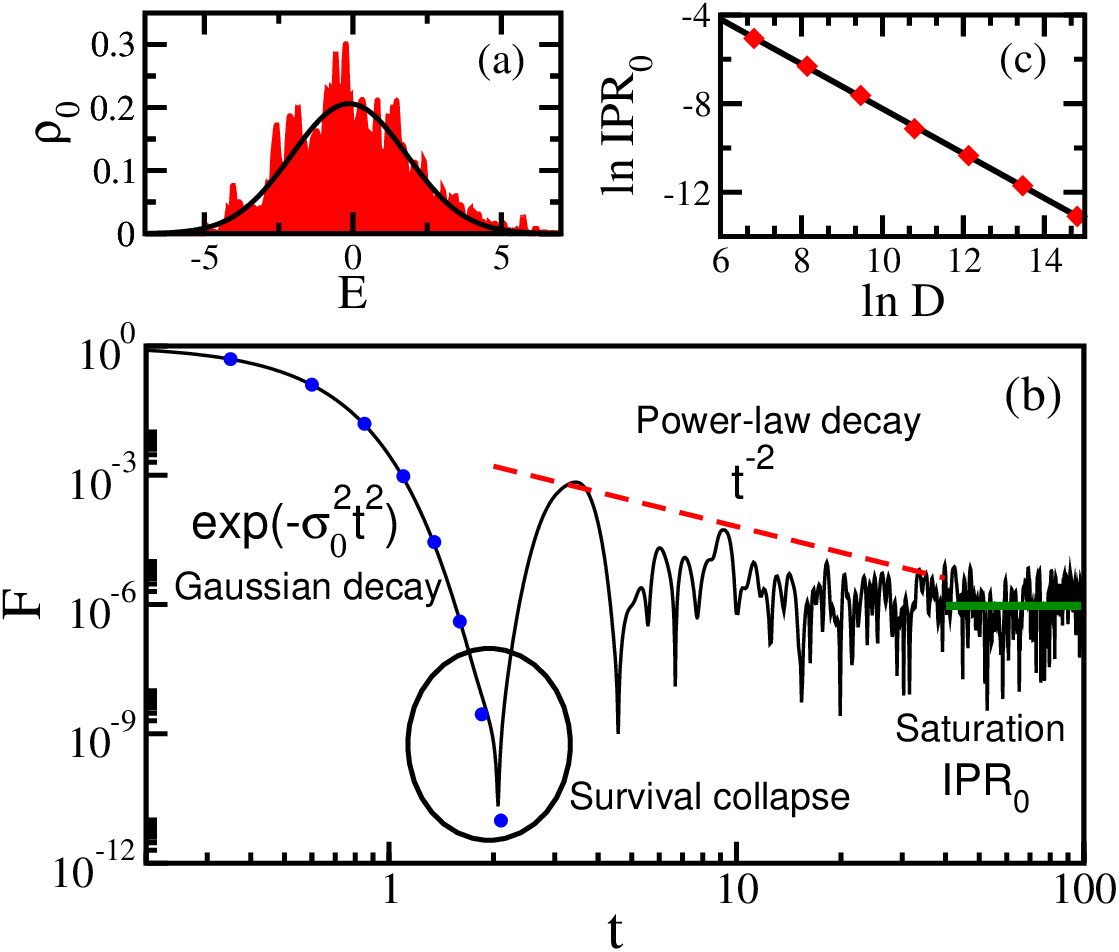}
\caption{Local density of states (a), survival probability decay (b), and scaling analysis of the $\text{IPR}_0$ (c)  for the N\'eel state evolving under $H$  (\ref{ham}) with $h=0$, $\Delta=1/2$, $\lambda=1$, and open boundaries. In (a): the shaded area is the numerical result and the solid line is a Gaussian with $E_0$ and $\sigma_0$ from Eq~(\ref{eq:NS}); $L=16$. In (b): the solid line is the numerical result obtained with EXPOKIT~\cite{Sidje1998,Expokit}, circles indicate the analytical Gaussian decay with $\sigma_0$ from Eq.~(\ref{eq:NS}), the dashed line is the time average coinciding with $t^{-2}$, and the thick horizontal line marks the saturation $\bar{F}=\text{IPR}_{0}$; $L=24$. In (c) the solid line is ${\rm IPR}_0 = 6/{\cal D}$. The first three points are obtained with exact diagonalization and the last four are infinite time averages computed with EXPOKIT.}
\label{fig:GaussNeel}  
\end{figure}

We start by investigating the survival probability of the N\'eel state evolving under the clean chaotic Hamiltonian (\ref{ham}) with $h=0$, $\Delta=1/2$, and $\lambda=1$. The perturbation that takes $H_0$  into this Hamiltonian is strong, since we need to change the anisotropy abruptly from $\Delta \rightarrow \infty$ to $\Delta=1/2$. As mentioned in Sec.~\ref{Sec:short}, the envelope of the LDOS should therefore have a Gaussian shape. This is confirmed in Fig.~\ref{fig:GaussNeel} (a). The Gaussian LDOS is nearly symmetric, since $E_0$ is close to the middle of the spectrum, and it agrees very well with the analytical envelope obtained with $E_0$ and $\sigma_0$ from Eq.~(\ref{eq:NS}). For initial states with $E_0$ closer to the edges of the spectrum, the LDOS acquires some degree of skewness~\cite{Torres2014PRAb}. 

Figure~\ref{fig:GaussNeel} (b) shows the evolution of the survival probability. Up to $t_P\sim2$, the decay is Gaussian, as anticipated from the Gaussian LDOS. The numerical curve agrees extremely well with the analytical expression using $\sigma_0$ from Eq~(\ref{eq:NS}). Interestingly,  in this initial decay, $F(t)$ reaches several orders of magnitude below the infinite time average $\bar{F}=\text{IPR}_{0}$. This pronounced dip has been referred to as survival collapse~\cite{Fiori2006,Fiori2009} and is further explained in Sec.~\ref{sec:collapse}.

For times longer than $t_P$, a power-law decay $\propto t^{-2}$ emerges. As mentioned before, this is expected to occur when the Gaussian LDOS is ergodically filled. This is indeed confimed with Fig.~\ref{fig:GaussNeel} (c). Using the values of $\text{IPR}_{0}$ for $L=12,14,16$ obtained from exact diagonalization [Eq.~(\ref{Eq:saturationF})], and the values for $L=18,20,22,24$ obtained from averages of the fluctuating values of $F(t)$ after saturation, we verify that $\text{IPR}_{0} \propto {\cal D}^{-1}$. For $L>16$, our computations are done with EXPOKIT~\cite{Sidje1998,Expokit}, which is a software package for the evolution of the matrix exponential $e^{-iHt}$ used when the Hamiltonian matrix is very large, but sparse.

The ergodic filling of the LDOS justifies Fourier transforming the continuous Gaussian function with the lower ($E_{low}$) and upper ($E_{up}$) bounds. This reveals the $t^{-2}$ decay of the survival probability,
\begin{equation}
F(t \gg \sigma _0^{ - 1})  \simeq \frac{1}{2\pi {\cal N}^2 \sigma_0^2 t^2}
\sum\limits_{k = up,low}  e^{ - (E_k - E_0)^2/\sigma_0^2}  .  
\label{decaylineMain}
\end{equation}
Above, $ {\cal N}$ is a normalization constant (see the derivation in the Appendix A).

The numerical curve for $F(t)$ at long times is affected by finite size effects, which cause the fluctuations observed in Fig.~\ref{fig:GaussNeel} (b). To smoothen the curve and substantiate the $t^{-2}$ behavior, we show with a dashed line the time-averaged survival probability defined as 
\begin{eqnarray}
C(t,t_0) \equiv \frac{1}{t-t_0}\int_{t_0}^t  F(\tau)d\tau.
\label{TA}
\end{eqnarray}
In practice, we actually average the logarithm of $F(t)$ instead of $F(t)$ via $(\ln t  - \ln t_0)^{-1}\!\int_{\ln t_0}^{\ln t}  \ln F (\eta )d\eta$. The power-law decay predominates after the survival collapse (see Sec.IV.A.1), so the values of $t_0$ and $t$ that we choose correspond, respectively, to the moment of the first revival of $F(t)$ and the time at which the survival probability saturates. Further support for the onset of the $t^{-2}$ behavior is  given in Sec.~\ref{furtherExamples} for other parameters and system sizes and in Sec.~\ref{sec:ftChaos} for a disordered chaotic Hamiltonian.

An estimate for the time $t_P$ where the algebraic decay starts can be obtained with the following approximation 
\[
e^{-\sigma_0^2 t_P^2} \sim F(t_P \gg \sigma_0^{-1}).
\]
It leads to
\[
{t_P} \simeq \sigma_0^{-1} \sqrt{ -W_{-1} \left(- \sum_{k=up,low} e^{-(E_k - E_0)^2/\sigma_0^2} /(2\pi) \right)},
\]
where $W$ is the Lambert $W$-function.

\subsubsection{Survival Collapse}
\label{sec:collapse}
The survival collapse is characterized by an abrupt drop of $F(t)$ by several orders of magnitude, which can bring it below the saturation point $\overline{F}=\mbox{IPR}_0$. This collapse can be understood as follows. Let us write the survival amplitude as a sum of two amplitudes, $A(t) = A_{\rm{G}}(t) + A_{\rm{R}}(t)$, so that 
\begin{eqnarray}
&&F(t) = |A_{\rm{G}}(t)|^2 + |A_{\rm{R}}(t)|^2 + A_{\rm{Int}}(t),
\label{eq:interference}
\\
&&A_{\rm{Int}}(t) = 2 \text{Re} \left[ A^*_{\rm{G}}(t) A_{\rm{R}}(t) \right]. \nonumber
\end{eqnarray} 
$A_{\rm{G}}(t)$ is obtained with the unbounded LDOS, 
\[
A_{\rm{G}}(t) = \int_{-\infty}^\infty  \rho_0 (E) e^{ - iEt}  dE .
\]
Its absolute square leads to a pure Gaussian decay when $\rho_0 (E)$ is Gaussian. $ A_{\rm{R}}(t) $ is the probability amplitude for the initial state to be reconstructed due to the presence of the bounds in the spectrum,
\[ 
A_{\rm{R}}(t) =  -\int_{-\infty}^{E_{low}}  \rho_0 (E) e^{ - iEt}  dE
-\int_{E_{up}}^{\infty}  \rho_0 (E) e^{ - iEt}  dE .
\]
$ A_{\rm{G}}(t)$ and $A_{\rm{R}}(t)$ can interfere destructively. When this happens, $A_{\rm{Int}}(t) <0$, which causes the low values of $F(t)$. 

\begin{figure}[htb]
\centering
\includegraphics*[width=3.3in]{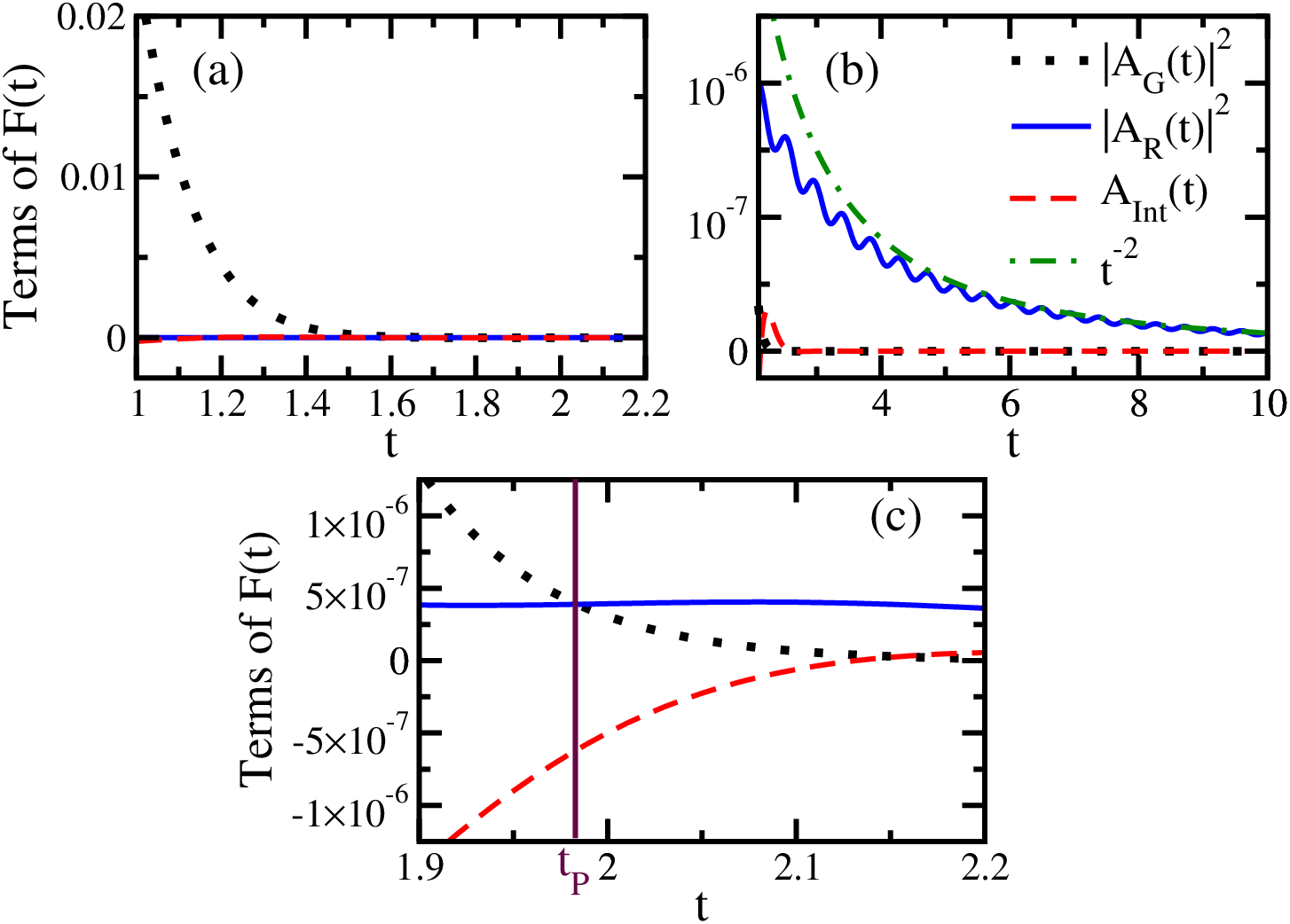}
\caption{$|A_{\rm{G}}(t)|^2$,  $|A_{\rm{R}}(t)|^2$, and the interference term $A_{\rm{Int}}(t)\equiv 2 \text{Re} \left[ A^*_{\rm{G}}(t) A_{\rm{R}}(t) \right]$  for different time scales: short times (a), long times (b), and the vicinity of $t_P$ (c). The dot-dashed line in (b) corresponds to $F(t) \propto t^{-2}$. The vertical solid line in (c) indicates $t_P\sim 1.98$. The data is obtained analytically for a Gaussian $\rho_0 (E)$ using the values of $E_0$ and $\sigma_0$ from Eq.~(\ref{eq:NS}) for a N\'eel state under $H$ with $h=0$, $\Delta=1/2$, $\lambda=1$, $L=16$. The values of $E_{low}$ and $E_{up}$ are obtained from exact diagonalization.}
\label{fig:collapse}  
\end{figure}

Figure~\ref{fig:collapse} shows $|A_{\rm{G}}(t)|^2$,  $|A_{\rm{R}}(t)|^2$, and $A_{\rm{Int}} (t)$ for an analytical Gaussian $\rho_0 (E)$ with lower and upper energy bounds obtained for a N\'eel state that evolves under $H$ with $h=0$, $\Delta=1/2$, $\lambda=1$, $L=16$. The Gaussian decay dominates the evolution when $t<2$ [Fig.~\ref{fig:collapse} (a)], the contributions from $|A_{\rm{R}}(t)|^2$ and $A_{\rm{Int}} (t)$ being negligible. In the contrast, the power-law behavior that emerges from $|A_{\rm{R}}(t)|^2$ controls the dynamics for $t>2$ [Fig.~\ref{fig:collapse} (b)]. 

The interference effect is significant at the crossover from the Gaussian to the power-law decay, where the contributions from $|A_{\rm{G}}(t_P)|^2 $ and $ |A_{\rm{R}}(t_P)|^2$ are similar. Given $E_0$, $\sigma_0$, $E_{low}$, and $E_{up}$, the crossover point $t_P$ can be obtained numerically from $ | A_{\rm{G}}(t)|^2 = |A_{\rm{R}} (t) |^2$. With the values used in Fig.~\ref{fig:collapse}, we find that $t_P \sim 1.98$. At the vicinity of $t_P$, the interference term $A_{\rm{Int}}(t)$ is negative and of absolute value similar to $|A_{\rm{G}}(t)|^2+|A_{\rm{R}}(t)|^2$,  as seen in Fig.~\ref{fig:collapse} (c). This is the region where the survival probability can be brought to very small values \cite{MugaPRA2006}.

\subsubsection{Further examples of $t^{-2}$ decays in clean systems}
\label{furtherExamples}

\begin{figure}[htb]
\centering
\includegraphics*[width=3.2in]{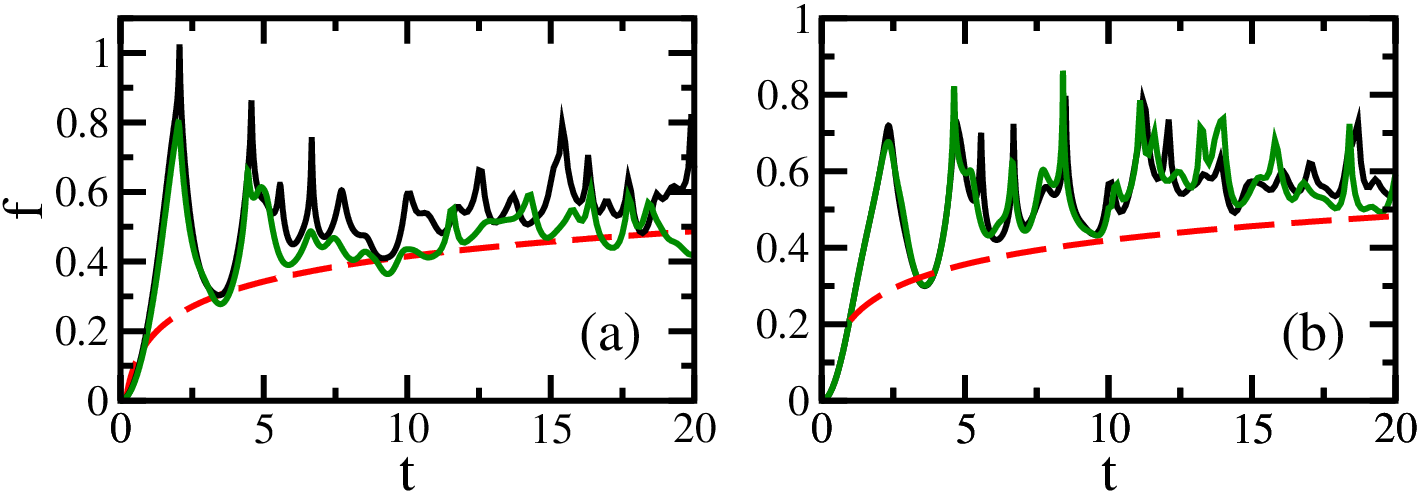}
\caption{Normalized survival probability $f(t)$ for the N\'eel state evolving under the chaotic $H$ (\ref{ham}) with $h=0$, $\lambda=1$, and open boundary conditions. In (a): $\Delta=1/2$ and in (b) $\Delta=0$. Light solid lines indicate $L=22$ and dark lines $L=24$. The dashed line is $ c -(1/L)\ln (t^{-2})$, where $c$ is a fitting constant.}
\label{fig:NeelDW}  
\end{figure}

A way to partially conceal the finite size effects is to consider the normalized survival probability used in~\cite{Andraschko2014},
\begin{equation}
f(t) =  - \frac{1}{L}\ln F(t) .
\label{smallf}
\end{equation}
This quantity is useful when comparing results for different system sizes. In Fig.~\ref{fig:NeelDW} we show $f(t)$ for the N\'eel state evolving under the chaotic Hamiltonian (\ref{ham}) with $h=0$ and $\lambda=1$ for systems with $L=22$ and $L=24$ and two values of the anisotropy parameter: (a) $\Delta=1/2$ and (b) $\Delta=0$. Both examples suggest that $F(t)\propto t^{-2}$.  Scaling analysis of $\text{IPR}_0$ for both cases  give $\text{IPR}_0\sim {\cal D}^{-1}$. 

\subsubsection{Disordered systems with $t^{-2}$ decay}
\label{sec:ftChaos}

\begin{figure}[htb]
\centering
\includegraphics*[width=3.2in]{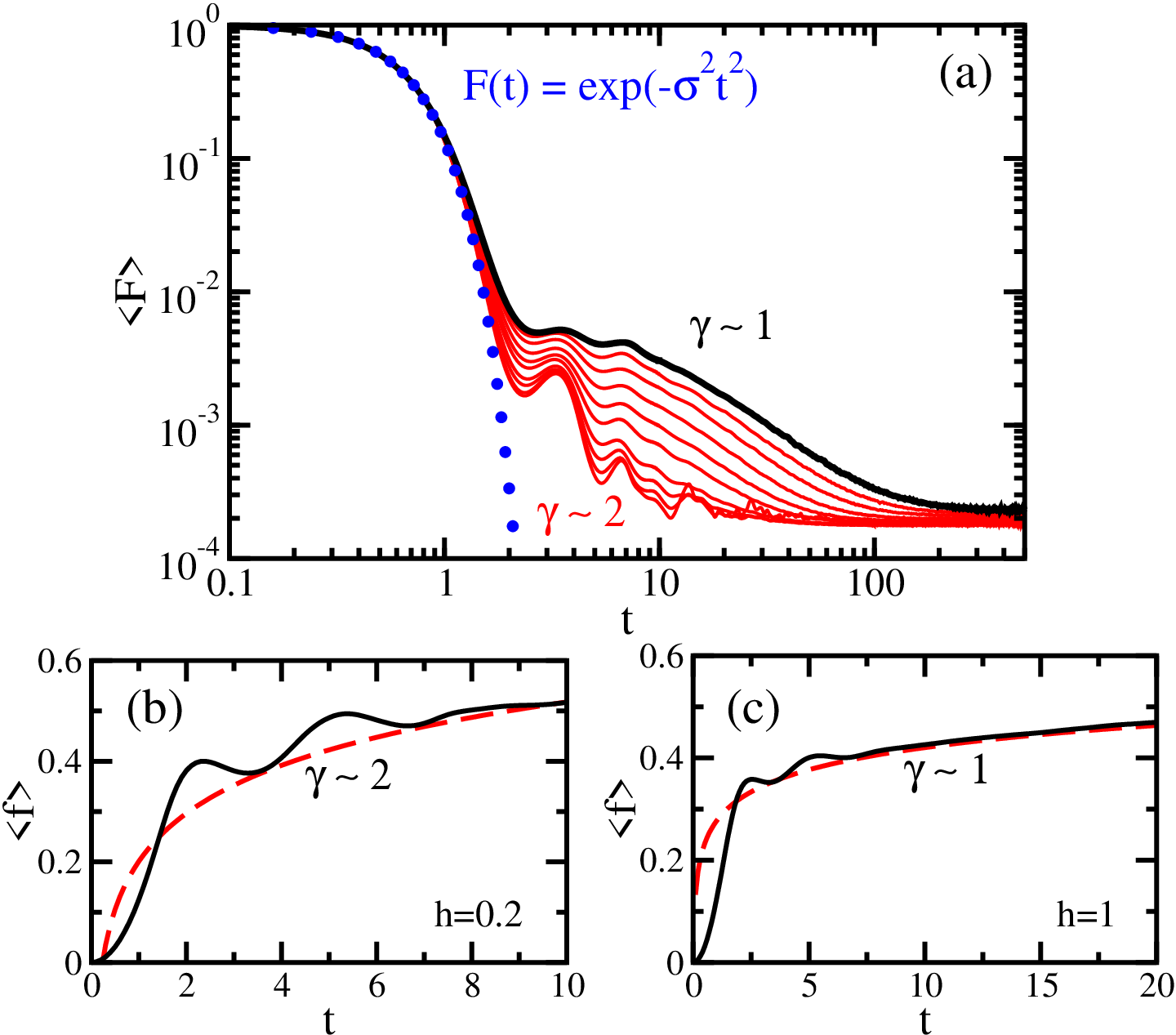}
\caption{Survival probability (a) and normalized survival probability (b), (c) for initial states corresponding to site-basis vectors and evolving according to $H$ (\ref{ham}) with $h$ in $[0.2,1]$, $\Delta=1$, $\lambda=0$, and closed boundaries. The average is performed over $10^5$ disorder realizations and initial states with energies $E_0$ close to the middle of the spectrum; $L=16$. In (a) the curves from bottom to top have $h=0.2, 0.3, \ldots 1$. The thick black line corresponds to $h=1$. In (b) and (c), the dashed lines are $ c -(1/L)\ln (t^{-\gamma})$, where $c$ is a fitting constant and $\gamma$ is indicated.}
\label{fig:disorderintermediary}  
\end{figure}

The $t^{-2}$ behavior is further reinforced by studying the dynamics under the disordered Hamiltonian (\ref{ham}) with $0<h<1$, $\Delta=1$, and $\lambda=0$. For these parameters, the Hamiltonian is chaotic. The initial states considered are site-basis vectors with $E_0$ away from the edges of the spectrum. In Fig.~\ref{fig:disorderintermediary} (a), we show the average of the survival probability, $\langle F(t) \rangle$, for different values of the disorder strength $\in [0.2,1]$. At intermediate times, the behavior is Gaussian. It is subsequently followed by power-law decays.

Due to the averages   over a total of $10^5$ data, including several realizations and initial states, the curves are smoother than those for the clean Hamiltonians in Figs.~\ref{fig:GaussNeel} and \ref{fig:NeelDW}. The average also erases the survival collapse.
 
The decay of the oscillations for the bottom curve in Fig.~\ref{fig:disorderintermediary} (a), which is obtained for $h=0.2$, follows a $t^{-2}$ behavior. This is made evident with the fitting line in Fig.~\ref{fig:disorderintermediary} (b), where this curve is isolated.

The value $\gamma=2$ is the limit for lattice many-body quantum systems with two-body interactions, as those described by $H$ (\ref{ham}). To increase the value of $\gamma$ above 2 and eventually reach the upper bound of $\gamma=3$ established by full random matrices [Eq.~(\ref{eq:FRMasymptotic})], one needs to increase the number of uncorrelated elements in the Hamiltonian matrix, so that the density of states and the LDOS will broaden and finally reach the semicircle shape. The intermediate values of $\gamma$ between $2$ and $3$  can be achieved with banded random matrices~\cite{TavoraARXIV}.

Banded random matrices were introduced in an attempt to better describe the details of real complex systems, where few particles interact simultaneously~\cite{Wigner1955}. Instead of having the matrix completely filled with random numbers, as in an full random matrix, the random numbers are restricted to a bandwidth around the diagonal. Beyond the band, the elements are either zero, as in Wigner banded random matrices~\cite{Wigner1955,Casati1996}, or very small, as in power-law banded random matrices~\cite{Mirlin1996}. By increasing the bandwidth from zero, one can cover all values of the power-law exponent, from 0 to 3. This was discussed and illustrated in Ref.~\cite{TavoraARXIV}.

We reiterate that power-law exponents $\gamma \geq 2$ reflect the ergodic filling of the LDOS. The algebraic decay in this case is caused by bounds in the spectrum and the initial state should eventually thermalize. 

The $t^{-2}$ decay was observed also in the interacting integrable $XXZ$ model with $\Delta=1$. The initial states considered were superpositions of equally weighted Bethe-ansatz eigenstates in a window of energy~\cite{DeguchiARXIV}. These initial states are constructed by choice to be ergodic. This may explain why, despite integrability, the exponent reaches the value $\gamma =2$.

\subsection{Power-law exponent $\pmb{1\le \gamma <2}$}
\label{sec:intermediate}

For the disordered Hamiltonian (\ref{ham}) with $\Delta=1$ and $\lambda=0$, level repulsion persists throughout the region of $h \in [0.2,1]$ and scaling analyses of the level of delocalization of the initial states written in the energy eigenbasis indicate that they are chaotic, $\text{IPR}_0 \propto {\cal D}^{-1}$ \cite{TorresARXIV}. The highest level of delocalization occurs for $h\sim 0.2$. As $h$ further increases, the level of delocalization decreases and so does $\gamma$, as seen in Fig.~\ref{fig:disorderintermediary} (a). The thick black line corresponds to $h=1$. This curve is also shown in Fig.~\ref{fig:disorderintermediary} (c)  together with the fitting line with $\gamma=1$. 

The cause for the power-law exponents $1 \leq \gamma <2 $ in these systems still needs to be understood. They suggest the existence of some minor correlations. The values of $\gamma$ could be a consequence of the interplay between these correlations and the energy bounds. Since the signatures of quantum chaos persist, we should still expect thermalization to take place.

We find exponents in this intermediate region also in integrable models. From the analytical expressions for the survival probability of the ground state evolving under the Ising model in a transverse field~\cite{Venuti2010} and under the $XY$ model~\cite{Venuti2011,Happola2012}, one can show that for long times $F(t) \propto \exp(L t^{-\gamma})$ with $\gamma=3/2$. Contrary to the disordered model above, the algebraic decay here develops only when $L t^{-\gamma} \ll 1$. It is possible that the nature of the power-law decay with $1 \leq \gamma <2 $ for the disordered model is different from what occurs for these integrable models.

\subsection{Power-law exponent $\pmb{\gamma< 1}$}
\label{XX}

In the disordered model, the power-law exponent becomes smaller than 1 when $h>1$. In this case, the LDOS is sparse and $\mbox{IPR}_0 \propto {\cal D}^{ - D_2 }$ with $D_2<1$.  In Refs.~\cite{Torres2015,Torres2015BJP,TorresARXIV}, we demonstrated that $D_2$ coincides with the value of the power-law exponent $\gamma$ of the survival probability decay. In this section, we show that also for the N\'eel state evolving under the clean noninteracting $XX$ model, $\gamma<1$ and it agrees with $D_2$. 

As shown in Ref.~\cite{Mazza2016}, only $2^{L/2}$ of the ${\cal D}$ overlaps $C_\alpha ^{(0)} = \langle \psi_{\alpha} |\text{NS} \rangle$ between the N\'eel state and the eigenstates of the $XX$ model are nonzero and their squared values are all the same,
\begin{equation}
|C_\alpha ^{(0)} |^2 = 2^{ - L/2}.
\label{eq:CalphaNS}
\end{equation}
[Details of the derivations are in the Appendix \ref{ap:XX}.] The LDOS is therefore very sparse. As a matter of fact, the ratio $\chi$ of the number of nonzero $|C_\alpha ^{(0)} |^2$ over the dimension of the Hilbert space goes exponentially to zero as $L\to\infty$, 
\begin{equation}
\chi=\frac{2^{L/2}}{{\cal D}} \xrightarrow{L \to \infty }  2^{ - (L + 1)/2} \, \sqrt{\pi L} .
\label{eq:sparsity}
\end{equation}

For open boundary conditions, the analytical expression for the survival probability is~\cite{Mazza2016}
\begin{equation}
F(t) = \prod\limits_{n = 1}^{L/2} {{{\cos }^2}} \left\{ {t\cos \left[ {\frac{{\pi n}}{{L + 1}}} \right]} \right\}.
\label{eq:FidNSopen}
\end{equation}
At long times, $L/\sqrt{t} \rightarrow 0$, choosing $L$ to be the largest length scale of the system, we find that the envelope of the decay of $F(t)$ is given by
\begin{equation}
F(\sqrt t  \gg L)\propto \exp\left( L t^{-1/2} \right)  \rightarrow   1 + L{t^{ - 1/2}}, 
\label{eq:FidNSxxlongtime}
\end{equation}
from where the power-law exponent is evident.

\begin{figure}[htb]
\centering
\hspace{-0.5 cm}
\includegraphics*[width=3.5in]{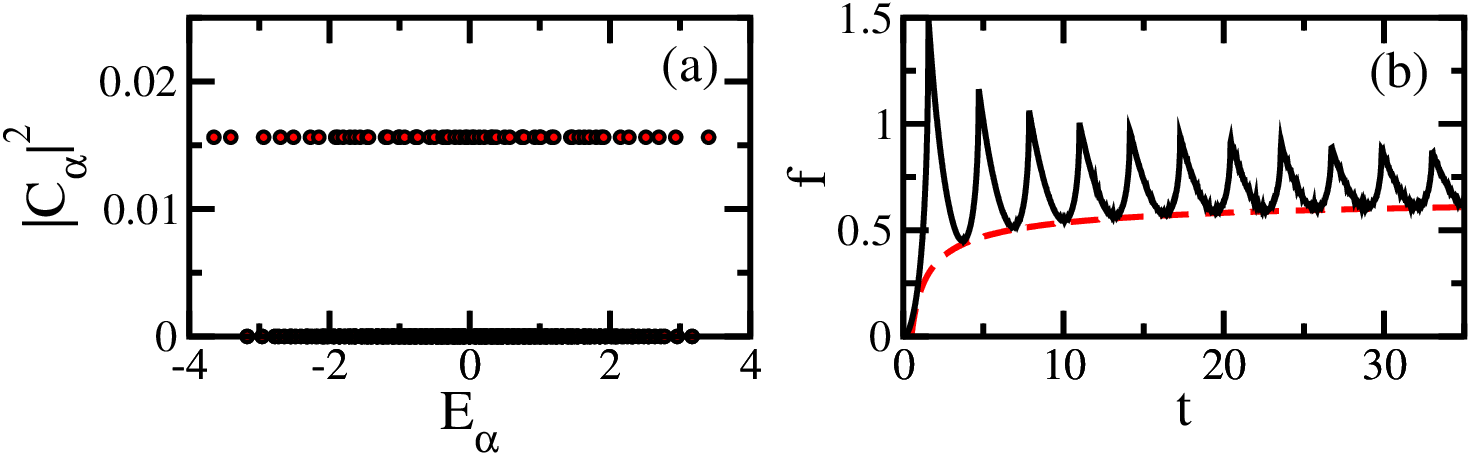}
\caption{Components $|C_\alpha ^{(0)}|^2$ (a) and $f(t)$ (b) for the N\'eel state evolving under the $XX$ Hamiltonian [Eq.~(\ref{ham}) with $h,\Delta,\lambda=0$, and open boundaries].  In (a):  $L=12$. In (b):  $f(t)$ (solid line) and $-(1/L) \ln (t^{-2})$ (dashed line); $L=2000$.}
\label{fig:fidXX}  
\end{figure}

In Fig.~\ref{fig:fidXX} (a), we give the values of $|C_\alpha ^{(0)} |^2$ as a function of the energies, making it clear that the number of nonzero components is small. Figure~\ref{fig:fidXX} (b) shows $f(t)$ from Eq.~(\ref{eq:FidNSopen}) and the $t^{-1/2}$ decay of the survival probability.

The power-law exponent $\gamma=1/2$ can also be derived from a scaling analysis, as done for systems with strong disorder. Using the fact that for the N\'eel state $\text{IPR}_0=2^{-L/2}$ and that from the Stirling approximation $\ln {\cal D} \simeq L\ln 2$, 
\begin{eqnarray}
\ln \text{IPR}_0 =  - D_2 \ln {\cal D}  \Rightarrow D_2 = \frac{1}{2} ,
\end{eqnarray}
which agrees with $\gamma$. Whether this relationship is a mere coincidence or is valid also for other initial states and integrable models remains to be elucidated. Similarly to the discussions in the end of Sec.~\ref{sec:intermediate}, we stress that the cause for the power-law decay in the integrable $XX$ model may not be exactly analogous to the one found in the disordered model. The source for the latter are the correlations in the eigenstates, measured equivalently with $D_2$ or $\gamma$. In the $XX$ model, the decay may be more involved, as  Eq.~(\ref{eq:FidNSxxlongtime}) suggests. 

Since the LDOS is sparse,  the initial state should not thermalize. This can be corroborated by comparing the diagonal entropy [Eq.~(\ref{d-entropy})] and the canonical entropy 
\begin{eqnarray}
S_{can} = \ln Z + E_0/T
\end{eqnarray}
where $Z = \sum\nolimits_{\alpha}  e^{ - E_{\alpha}/T}$ is the partition function, $T$ is the temperature, and the Boltzmann constant is set to 1. From Eq.~(\ref{eq:CalphaNS}),
\begin{equation}
S_d = \frac{L}{2}\ln 2 .
\label{eq:SdNS}
\end{equation}
For the N\'eel state in the $XX$ Hamiltonian, $E_0=0$, so the temperature is infinite and $e^{ - E_{\alpha }/T} \to 1$, so  $Z = \cal D$. The thermal entropy for large $L$ is therefore
\begin{eqnarray}
S_{can}  \simeq L\ln 2 .
\end{eqnarray}
The fact that $S_{can}$ and $S_d$ do not coincide implies lack of thermalization. 

We note that care should be taken when computing the diagonal entropy. The expression (\ref{d-entropy}) is appropriate for systems without too many degeneracies.

\subsection{A special case}
\label{DW}
The domain wall state evolving under the $XX$ model is a very special case, where the power-law decay does not seem to develop~\cite{Mazza2016}. As seen from Eq.~(\ref{eq:DW}), the width of the LDOS does not depend on the system size, it is fixed at $\sigma_0=J/2$. In addition, for the $XX$ model, $E_0 =0$. Therefore, as $L$ increases, the LDOS becomes an increasingly better filled Gaussian,  $\rho _0(E) = e^{ - 2 E^2}/\sqrt {\pi /2}$. The contributions from the tails become less and less relevant and the survival probability decay approaches a perfect Gaussian behavior, $F(t)=\exp(-{t^2}/4)$.

\begin{figure}[htb]
\centering
\includegraphics*[width=3.2in]{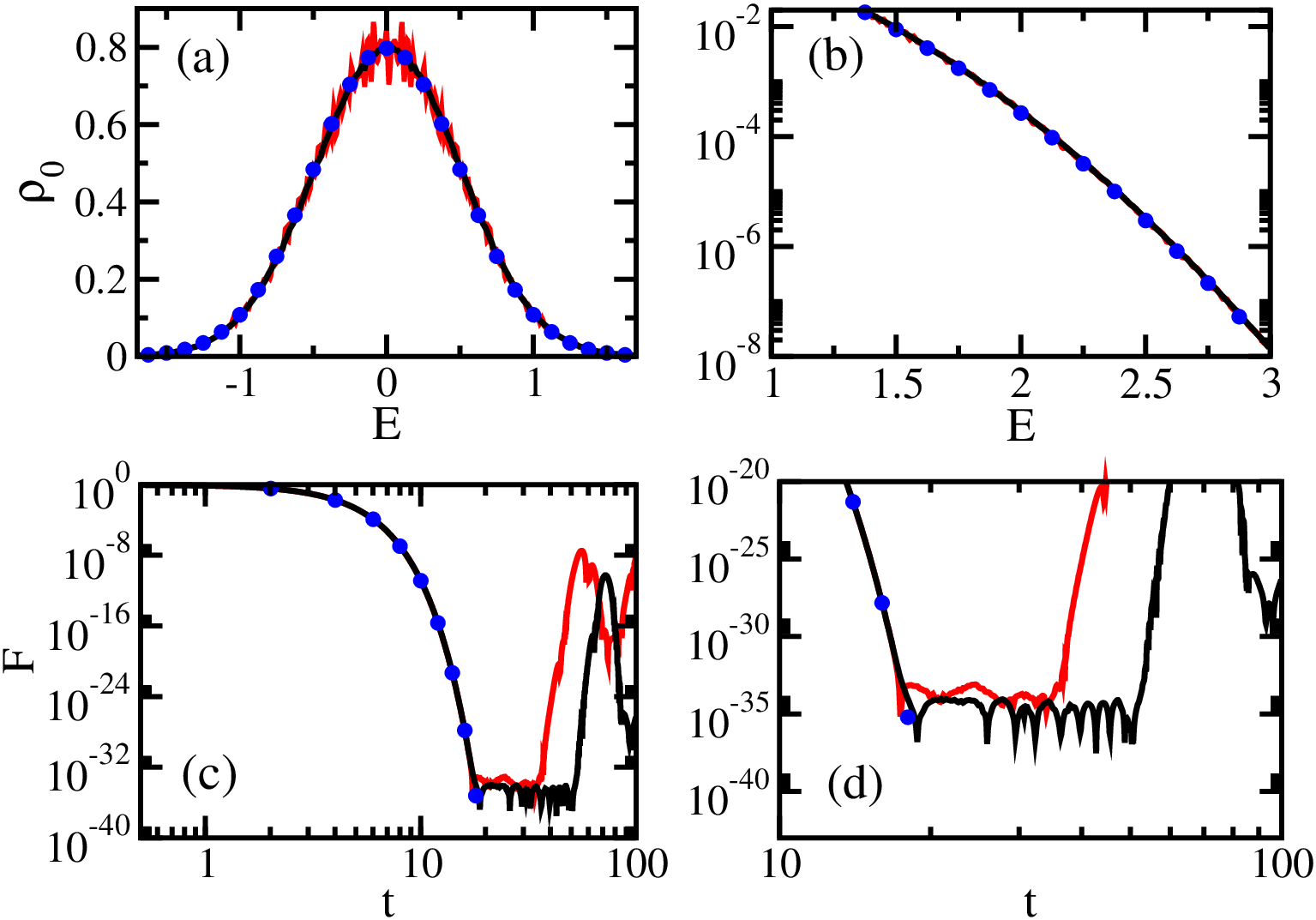}
\caption{LDOS (a),(b) and survival probability [Eq.(\ref{equationDWopen})]; (c),(d) for the domain wall state evolving under the $XX$ Hamiltonian [Eq.~(\ref{ham}) with $h,\Delta,\lambda=0$, and open boundaries]. Two system sizes are shown in all panels. The (red) light lines indicate $L=24$; the (black) dark lines $L=30$; and the (blue) circles the Gaussian curves ${\rho _0}(E) = {e^{ - 2{E^2}}}/\sqrt {\pi /2}$ and $F(t) = \exp ( - {t^2}/4)$.}
\label{fig:fidXXdw}  
\end{figure}

In Figs.~\ref{fig:fidXXdw} (a) and (b), we confirm that the LDOS width remains unchanged as $L$ increases. Two system sizes, $L=24$ and $30$, are considered. Figure~\ref{fig:fidXXdw} (b), in particular, emphasizes the negligible contributions from the tails already at energies smaller than the energy bounds of the entire spectrum [cf. Fig.~\ref{fig:fidXXdw} (b) and Fig.~\ref{fig:fidXX} (a)].

For open boundary conditions, the exact expression for the survival probability is 
\begin{eqnarray}
F(t) &=&\left( \frac{2^{L - 1}}{L + 1} \right)^L\left| \sum\limits_{\{ k_n\} } \left\{ \cos \left[ t\sum\limits_{n = 1}^{L/2} \cos  \,{k_n} \right]\prod\limits_{n = 1}^{L/2} \sin^2 k_n  \right. \right.\nonumber\\
&\times& \!\!\! \left. \left.  \prod\limits_{m = n + 1}^{L/2}  \!\! \sin^2 \left( \frac{k_n - k_m}{2} \right) \sin^2 \left( \frac{k_n + k_m}{2} \right) \right\} \right|^2.
\label{equationDWopen}
\end{eqnarray}
[Details on how to obtain Eq.~(\ref{equationDWopen}) are shown in the appendix~\ref{ap:DW}.] The decay obtained with this expression is compared in Figs.~\ref{fig:fidXXdw} (c) and (d) with $F(t)=\exp(-{t^2}/4)$. The agreement is extremely good. After $t \simeq {\cal O}(L)$, an oscillatory behavior sets in. When that happens, $F(t)$ is already essentially zero, so if a decay rate still exists, it is very difficult to estimate. Furthermore, as seen in Fig.~\ref{fig:fidXXdw} (d), the value of $F(t\simeq L)$ goes to smaller numbers as $L$ increases and it stays there for longer periods of time. The revivals observed at later times occur when the excitations eventually reach the system's boundary. In the thermodynamic limit, we should therefore expect $F(t)$ to decay to zero and recurrences to be nonexistent.


\section{Conclusions} 

We investigated the long-time decay of the survival probability in isolated lattice many-body quantum systems. We considered integrable and chaotic, interacting and noninteracting, and clean and disordered systems. Our results showed that  for all of these systems the long-time decay is  algebraic, $F(t)\propto t^{-\gamma}$.

There is a clear-cut relationship between the power-law decay exponent $\gamma$ and the degree of delocalization of the initial state written in the energy eigenbasis. For a maximally delocalized initial state, its weighted energy distribution (LDOS) is ergodically filled and the power-law decay is caused by the ever-present bounds in the spectrum. For realistic models with two-body interactions, this leads to $\gamma=2$. When the initial state is no longer chaotic, so that $\text{IPR}_0 \propto {\cal D}^{-D_2}$ with $D_2<1$, then $\gamma =D_2$. 

Since ergodicity guarantees thermalization, we were able to establish a criterion for thermalization based purely on the dynamics of the system at long times. This is a significant result, because various experimental studies of many-body systems focus on time evolutions. We can summarize our main findings as follows:
\begin{eqnarray}
& & \hspace{0.3 cm}\text{nonchaotic initial state, sparse LDOS,}   \nonumber \\
& \gamma<1  & \hspace{0.3 cm} \text{correlated eigenstates.}   \nonumber \\
\nonumber \\
& & \hspace{0.3 cm}\text{chaotic initial state, ergodically filled LDOS,}   \nonumber \\
& \gamma=2  & \hspace{0.3 cm} \text{Wigner-Dyson level repulsion,}    \nonumber \\
& & \hspace{0.3 cm} \text{uncorrelated eigenstates, thermalization.}   \nonumber  
\end{eqnarray}

Still an open question is the case of the intermediate values of $\gamma \in [1,2)$. They may be due to a competition between minor correlations and energy bounds.


\begin{acknowledgments}
This work was supported by the NSF grant No.~DMR-1147430. EJTH acknowledges funding from CONACYT, PRODEP-SEP, and VIEP-BUAP, Mexico. 
\end{acknowledgments}

\appendix

\section{Power-law Exponents for Absolutely Integrable LDOS}
\label{RigRes}

As mentioned in Sec.~\ref{subcase1}, an absolutely integrable $\rho_0(E)$ with a Gaussian or Lorentzian shape falls into the category of  Case 1 (i), where for long times $F(t) \propto t^{-2}$.  For Case 1 (ii),  the decay of the survival probability is faster, having $\gamma >2$. It holds when the LDOS is a function that goes to zero at the energy bound.

\subsection{ Case 1 (i): Gaussian LDOS}
The survival amplitude for the Gaussian LDOS with a lower bound $E_{low}$ is given by, 
\begin{eqnarray}
A(t) &=&  \frac{1}{\sqrt {2\pi \sigma_0^2} }\int_{E_{low}}^\infty  dE\, e^{ - iEt} e^{ -( E - E_0)^2/2\sigma_0^2} \nonumber \\
&=& \frac{e^{-i E_{low} t}  }{\sqrt {2\pi \sigma_0^2} } \int_{0}^\infty  d{\cal E} \, e^{ - i{\cal E} t} e^{ -( {\cal E}  + E_{low} - E_0)^2/2\sigma_0^2} \nonumber
\end{eqnarray}
where ${\cal E}  = E - E_{low}$
For long times, the  first exponential inside the integral
oscillates very fast, unless ${\cal E} $ is very small. If we then set ${\cal E} =0$ in the second exponential inside the integral, we find that
\[
\int_{0}^\infty  d{\cal E} \, e^{ - i{\cal E} t} e^{ -( E_{low} - E_0)^2/2\sigma_0^2} \propto t^{-1}
\]
so
\[
F(t) = |A(t)|^2 \propto t^{-2}.
\]

A more rigorous way to obtain $F(t)$ at long times takes into account the lower bound $E_{low}$ and the upper bound $E_{up}$ as, 
\begin{equation}
A(t) =  \frac{1}{{\cal N} \sqrt {2\pi \sigma_0^2} }\int_{E_{low}}^{E_{up}}  dE\, e^{ - iEt} e^{ -( E - E_0)^2/2\sigma_0^2},
\end{equation}
where $\cal N$ is the normalization constant,
\begin{equation}
{\cal N} = \frac{1}{2} \left[ {\rm erf} \left( \frac{E_0 - E_{low} }{\sqrt{2\sigma_0^2} } \right) - {\rm erf} \left( \frac{E_0 - E_{up}}{\sqrt {2\sigma_0^2} } \right) \right]
\end{equation}
and $\text{erf}$ is the error function.  The survival amplitude can be obtained analytically and reads
\begin{align}
A(t) = \frac{1}{2 {\cal N}}  e^{ - \sigma_0^2 t^2/2 + i E_0 t}  &\left[   {\rm erf}  \left( \frac{E_0 - E_{low} + i\sigma_0^2t}{\sqrt{2} \sigma_0 } \right) \right. \nonumber \\
& -   {\rm erf}   \left.  \left( \frac{E_0 - E_{up} + i\sigma_0^2t}{\sqrt{2} \sigma_0 } \right) \right] .\nonumber
\end{align}
The corresponding survival probability is then,
\begin{align}
F(t) = \frac{ e^{ - \sigma_0^2 t^2 } }{4 {\cal N}^2}    & \left| \left[   {\rm erf}  \left( \frac{E_0 - E_{low} + i\sigma_0^2t}{\sqrt{2} \sigma_0 } \right) \right. \right. \nonumber \\
& -   {\rm erf}   \left. \left.  \left( \frac{E_0 - E_{up} + i\sigma_0^2t}{\sqrt{2} \sigma_0 } \right) \right] \right|^2 .
\end{align}
In the limit $t \gg 1/\sigma_0$, 
\begin{eqnarray}
&&F(t \gg \sigma _0^{ - 1})\nonumber\\
&& \simeq \frac{1}{2\pi {\cal N}^2 \sigma_0^2 t^2} \left[ e^{ - (E_{up} - E_0)^2/\sigma_0^2} +e^{ - (E_{low} - E_0)^2/\sigma_0^2} \right.\nonumber\\
&&\left.  - 2 e^{ - \left( E_{up}^2 - 2 E_0 (E_{low} + E_{up}) + E_{low}^2 + 2E_0^2 \right)/2\sigma_0^2}\cos (\Delta Et) \right] , \nonumber
\end{eqnarray}
where $\Delta E = E_{up} - E_{low}$ is the width of the spectrum.  Averaging out the oscillations from the cosine term, this expression becomes
\begin{equation}
F(t \gg \sigma _0^{ - 1})  \simeq \frac{1}{2\pi {\cal N}^2 \sigma_0^2 t^2}
\sum\limits_{k = up,low}  e^{ - (E_k - E_0)^2/\sigma_0^2}  ,  
\label{decayline}
\end{equation}
from where the decay $\propto t^{-2}$ is evident.

\subsection{ Case 1 (i): Lorentzian LDOS}
In the case of a Lorentzian LDOS, we have
\begin{eqnarray}
A(t) &=& \int_{E_{low}}^{\infty} \frac{1}{2\pi} \frac{\Gamma_0}{(E_0 - E)^2 +
 \Gamma_0^2 /4} e^{- i E t} dE , \nonumber \\ [10pt]
 &=&\frac{\Gamma_0 e^{-i E_{low} t}}{2\pi}   \int_{0}^{\infty} f({\cal E}) e^{- i {\cal E} t} d{\cal E} \nonumber \\
 \text{where} &&f({\cal E}) =  \frac{1}{(E_0 - {\cal E} - E_{low})^2 +
 \Gamma_0^2 /4}  , \nonumber
\end{eqnarray}
  ${\cal E}  = E - E_{low}$, and $\Gamma_0$ is the width of the distribution.
The integral above, can be solved by replacing it with a contour integral in the complex plane~\cite{MugaBook}. The complex contour has three parts, the positive real energy axis from zero to $\infty$, the arc of infinite radius running clockwise from the positive real axis to the negative imaginary axis, and the negative imaginary axis going from $-i\infty$ to the origin,
\begin{eqnarray}
 \oint\limits_{\cal C} f({\cal E}) e^{- i {\cal E} t} d{\cal E} &=& \int_0^\infty  f({\cal E}) e^{- i {\cal E} t} d{\cal E} \nonumber \\
 &+& \int_{arc}  f({\cal E}) e^{- i {\cal E} t} d{\cal E} \nonumber \\
 &+& \int_{-i\infty}^0  f({\cal E}) e^{- i {\cal E} t} d{\cal E}.
 \label{eq:contour}
\end{eqnarray}
As it is often the case, the integration along the arc vanishes. Using ${\cal E} = -i \varepsilon$, we are left with
\begin{eqnarray}
I(t) &\equiv& \int_0^\infty  f({\cal E}) e^{- i {\cal E} t} d{\cal E} \nonumber \\
&=&  \oint\limits_{\cal C} f({\cal E}) e^{- i {\cal E} t} d{\cal E} + \int_0^{-i\infty}  f({\cal E}) e^{- i {\cal E} t} d{\cal E} \nonumber \\
&=& \oint\limits_{\cal C} f({\cal E}) e^{- i {\cal E} t} d{\cal E} - i\int_0^\infty  f(- i\varepsilon) \, e^{ - \varepsilon t} d\varepsilon \nonumber \\
&=& I_1(t) + I_2(t) . \nonumber
\end{eqnarray}
The contour integral above is solved with residues. It has a pole at ${\cal E} = E_0 - E_{low} -i \Gamma_0/2$, which leads to the exponential decay,
\begin{eqnarray}
I_1(t) &=&\oint \frac{ 
\frac{e^{- i {\cal E} t}}{ [{\cal E}- (E_0 - E_{low}) ] - i \Gamma_0/2} }{ [{\cal E} - (E_0 - E_{low}) ] + i \Gamma_0/2} \nonumber \\[10 pt]
&&  = (- 2\pi i)  \frac{e^{-i (E_0 - E_{low}) t}  e^{- i (-i \Gamma_0 /2) t}}{  - i  \Gamma_0 } \propto e^{- \frac{\Gamma_0 t}{2}}.
\nonumber
\label{expPole1}
\end{eqnarray}
The second integral leads to the power-law decay. Since the integrand goes to zero for long times unless ${\varepsilon}$ is small, we set ${\varepsilon}=0$ in $f(- i\varepsilon)$,
\[
I_2(t) = - i \int_0^\infty \frac{e^{ - \varepsilon t}}{(E_0 - E_{low})^2 +
 \Gamma_0^2 /4} d\varepsilon \propto t^{-1}.
 \]
Just as for the Gaussian, $I_2(t)$ leads to $F(t) \propto t^{-2}$.

\subsection{Case 1 (i): Gaussian LDOS with exponential tails}
\label{ap:Etails}
Studies of the nuclear shell model have dealt with an LDOS that is Gaussian in the center and has exponential tails~\cite{Frazier1996}. To obtain the power-law decay exponent $\gamma$ for this case, we shift the LDOS and set the lower energy bound ${{E_{low}}}=0$. At long times, the relevant part of the LDOS is that where $\rho _0(E \simeq 0)$ and the exponential tail becomes dominant. This holds for a certain energy scale $[0, {\tilde E}]$, for which we can write
\begin{eqnarray}
A(t \to \infty ) &\simeq& \int_0^{\tilde E} dE e^{ - iEt} \rho _0(E \simeq 0) \nonumber\\
\,\,\,\,\,\,\,\,\,\,\,\,\,\,\,\,\,\,\,\,\,\, &\simeq& \int_0^{\tilde E} dE\,e^{ - iEt} e^{E} \nonumber\\
\,\,\,\,\,\,\,\,\,\,\,\,\,\,\,\,\,\,\,\,\,\, &\simeq& \frac{{({e^{- \tilde E( it-1)}} - 1)}}{(-i t+1)}.
\end{eqnarray}
The power-law decay of $F(t)$ is then $\propto {t^{ - 2}}$.

\subsection{Case 1 (ii)}

The derivation of Eq.~(\ref{eq:xi_gamma}) was done rigorously in Ref.~\cite{Erdelyi1956,Urbanowski2009}. Here, we provide a less rigorous alternative that incorporates Case 1 (i) and (ii) in a single equation. It was proposed in Ref.~\cite{Nowakowski2008} and goes as follows. Suppose that $\rho_0(E)$ has the following structure,
\[
\rho _0(E) = (E - E_{low})^\xi P(E)\eta (E)\Theta(E-E_{low})
\]
where now $\xi \ge 0$, $P(E)$ may contain poles, $\eta (E)$ is an analytical function with $\eta (E \to \infty ) \to 0$, and $\Theta(E-E_{low})$ is the Heaviside step function.
The survival amplitude then reads,
\[
A(t) = \int_{E_{low}}^\infty  dE\,{\rm e }^{ - iEt}(E - E_{low})^\xi P(E)\eta (E) .
\]
As done for the Lorentzian LDOS above, it is convenient to write this integral in the complex plane. It also helps to shift the lower bound to the origin of the complex plane by defining ${\cal E}  = E - {E_{low}}$,
\begin{eqnarray}
&&A(t)  = e^{ - iE_{low}t} \int_0^\infty  {d{\cal E}  \,{{\rm{e}}^{ - i{\cal E}  t}}{{\cal E}^\xi }P({\cal E}  + {E_{low}})\eta ({\cal E}   + {E_{low}})} \nonumber\\
&&\,\,\,\,\,\,\,\,\,\, \equiv e^{ - iE_{low}t} I(t) . \nonumber
\end{eqnarray}
The complex contour is the same used in Eq.~(\ref{eq:contour}) and again the integral along the arc is assumed to vanish, so
\begin{eqnarray}
&&I(t) = \oint_C {d{\cal E} \,{{\rm{e}}^{ - i{\cal E} t}}{{\cal E} ^\xi }P({\cal E}  + {E_{low}})\eta ({\cal E}  + {E_{low}})} \nonumber\\
&&\,\,\,\, + {( - i)^{\xi  + 1}}\int_0^\infty  {d\varepsilon \,{{\rm{e}}^{ - \varepsilon t}}{\varepsilon ^\xi }P({E_{low}} - i\varepsilon )\eta ({E_{low}} - i\varepsilon )} , \nonumber
\label{Eupinfinite}
\end{eqnarray}
where in the second integral we used ${\cal E} = -i \varepsilon$. Similarly to what we saw for the Lorentzian LDOS, the first integral depends on the poles  of $P(E)$ and it leads to very fast decays. It is the second integral that leads to much slower decays and therefore dominates the behavior of $F(t)$ at long times. Since  for large $t$ only small values of $\varepsilon$ contribute to the second integral, we set $\varepsilon=0$ in $P$ and $\eta$ and obtain
\begin{eqnarray}
&&I(t \to \infty ) \simeq {\cal C}{t^{ - \xi  - 1}} , \nonumber \\[5 pt]
\text{where} && {\cal C} = {( - i)^{\xi  + 1}}P({E_{low}})\eta ({E_{low}})\Gamma (\xi  + 1) . \nonumber
\end{eqnarray}
Hence, the survival probability decays as
\begin{equation}
F(t \to \infty ) \simeq |{\cal C}{|^2}{t^{ - 2(\xi  + 1)}},
\end{equation}
which agrees with Eq.~(\ref{eq:xi_gamma}), but includes also $\xi=0$.
This result remains valid if we include a finite energy upper bound $E_{up}$, since the behavior at long times is controlled by the lower spectrum bound.

\section{N\'eel initial state in the $XX$ Model}
\label{ap:XX}

Using the Jordan-Wigner transformation~\cite{Jordan1928}, the Hamiltonian for the noninteracting $XX$ chain ($h,\Delta, \lambda=0$) becomes
\begin{equation}
H_{XX} = \sum\limits_{k_n} \epsilon _{k_n} \, c_{k_n}^\dag c_{k_n} ,\hspace{0.9 cm } \epsilon_{k_n} = \cos k_n .
\end{equation}
The values of the momenta $k_n$ depend on the boundary conditions.

\subsection{Open boundary conditions}

Following Ref.~\cite{Mazza2016},  the momenta $k_n$ for open boundary conditions are
\begin{equation}
k_n \in {\cal K} = \left\{ {\frac{{\pi n}}{{L + 1}}} \right\}_{n = 1}^L
\label{possiblemomentaopen}
\end{equation}
The eigenstates of the ${\cal S}^z=0$ sector are
\begin{equation}
| \psi_{\alpha} \rangle  = \prod\limits_{n  = 1}^{L/2} c_{k_n }^\dag  | 0 \rangle ,
\end{equation}
where $\alpha  \equiv {\{ {k_n}\} _{n = 1,...,L/2}}$ and $| 0 \rangle$ is the vacuum state (all spins pointing down in the $z$-direction).
The eigenvalues are obtained with subsets $ \{ {k_n}\}$ of $\cal K$,
\begin{eqnarray}
E_{\alpha} = \sum\nolimits_{\{ k_n\}  \in {\cal K}} {\cos (k_n)} .
\label{eq:eigenvaluesXX}
\end{eqnarray}

To clarify the notation, we consider the case where $L=4$. For this choice, the set ${\cal K}$ is
\begin{eqnarray}
{\cal K} = \left\{ {\frac{\pi }{5},\frac{{2\pi }}{5},\frac{{3\pi }}{5},\frac{{4\pi }}{5}} \right\}
\end{eqnarray}
and the subsets $\alpha = \{ {k_1},{k_2}\}$ belong to
\begin{eqnarray}
&&\alpha  \in \left\{ {\left\{ {\frac{\pi }{5},\frac{{2\pi }}{5}} \right\},\left\{ {\frac{\pi }{5},\frac{{3\pi }}{5}} \right\},\left\{ {\frac{\pi }{5},\frac{{4\pi }}{5}} \right\}} \right.,\nonumber\\
&&\left. {\,\,\,\,\,\,\,\,\,\left\{ {\frac{{2\pi }}{5},\frac{{3\pi }}{5}} \right\},\left\{ {\frac{{2\pi }}{5},\frac{{4\pi }}{5}} \right\},\left\{ {\frac{{3\pi }}{5},\frac{{4\pi }}{5}} \right\}} \right\} .
\end{eqnarray}

The N\'eel state  can be written in terms of fermionic operators as
\begin{equation}
| {\rm{NS}} \rangle= \prod\limits_{j = 1}^{L/2} c_{2j - 1}^\dag | 0 \rangle  ,
\end{equation}
where the $c_{j}$'s are the inverse Fourier transform of the operators ${c_{{k_n}}}$,
\begin{eqnarray}
c_{{k_n}}^\dag  = \sqrt {\frac{2}{{L + 1}}} \sum\limits_{j = 1}^L {\sin ({k_n}j)} c_j^\dag .
\label{inverseFT}
\end{eqnarray}
The overlaps 
\begin{eqnarray}
C_\alpha ^{(0)} = \langle {\psi _\alpha }|{\rm{NS}}\rangle  = \langle 0 |\prod\limits_{n,j  = 1}^{L/2} {{c_{{k_n }}}c_{2j - 1}^\dag \left| 0 \right\rangle }
\label{olNS}
\end{eqnarray}
can be expressed in terms of Slater determinants applying Wick's theorem,
\begin{eqnarray}
C_\alpha ^{(0)} = {\left( {\frac{2}{{L + 1}}} \right)^{\frac{L}{4}}}\mathop {\det }\limits_{1 \le j,n  \le L/2} \sin [(2j-1){k_n }] .
\label{semnome}
\end{eqnarray}

Considering again the example above, for $L=4$ and choosing $\alpha = \{ {k_1},{k_2}\}  = \left\{ {\pi /5,2\pi /5} \right\}$, the overlap is
\begin{eqnarray}
&&C_{\left\{ {\pi /5,2\pi /5} \right\}}^{(0)} = \frac{2}{5}{\left( {\frac{1}{{\sqrt 8 }}} \right)^2}(5 - \sqrt 5 )\nonumber\\
&&\,\,\,\,\,\,\, \times \det \left( {\begin{array}{*{20}{c}}
1&{(1 + \sqrt 5 )/2}\\
{(1 + \sqrt 5 )/2}&{ - 1}
\end{array}} \right)
\end{eqnarray}

From the overlaps (\ref{semnome}), Eq.~(\ref{eq:FidNSopen}) in the main text can be obtained as done in \cite{Mazza2016}.

\subsection{Periodic boundary conditions}

For periodic boundary conditions, the inverse Fourier transform of the operators $c_{j}$'s are:
\begin{eqnarray}
c_{{k_n}}^\dag  = \sqrt {\frac{1}{L}} \sum\limits_{j = 1}^L {{{\rm{e}}^{{ik_n}j}}} c_j^\dag 
\label{inverseFTclosed}
\end{eqnarray}
where
\begin{equation}
k_n \in {\cal K} = \left\{ {\frac{{(2n - a)\pi }}{L}} \right\}_{n = 1}^L
\label{possiblemomentaclosed}
\end{equation}
and $a=0$ $(1)$ if $L/2$ is odd (even). The corresponding overlaps~(\ref{olNS}) read~\cite{Mazza2016}:
\begin{equation}
C_\alpha ^{(0)} = \left( \frac{1}{L} \right)^{L/4} e^{  i\sum\nolimits_{n = 1}^{L/2} k_n }\mathop {\det }\limits_{1 \le j,n \le L/2} \left( e^{-2i k_n j} \right),
\end{equation}
and the survival probability is given by~\cite{Andraschko2014,Mazza2016},
\begin{equation}
F(t) = \prod_{n = 1}^{L/2} \cos^2 \left\{ t \cos \left[ \frac{(2n - a)\pi }{L} \right] \right\}  .
\label{eq:FidNS}
\end{equation}
At long times, we find the same decay as in Eq.~(\ref{eq:FidNSxxlongtime}).

\section{Domain wall state in the $XX$ Model}
\label{ap:DW}
The domain wall can be written as
\begin{eqnarray}
\left| {{\rm{DW}}} \right\rangle  = \prod\limits_{j = 1}^{L/2} {c_j^\dag \left| 0 \right\rangle } .
\end{eqnarray}
Below, we show how we obtain the overlaps and the expression for the survival probability for open and periodic boundary conditions.

\subsection{Open boundary conditions}

For open boundary conditions, the overlaps $C_\alpha ^{(0)}$ are
\begin{eqnarray}
C_\alpha ^{(0)} = \langle {\psi _\alpha }|{\rm{DW}}\rangle  = \langle 0|\prod\limits_{n,j  = 1}^{L/2} {{c_{{k_n }}}c_j^\dag \left| 0 \right\rangle } .
\end{eqnarray}
Using Eq.~(\ref{inverseFT}) and applying Wick's theorem, this expression becomes
\begin{eqnarray}
C_\alpha ^{(0)} = {\left( {\frac{2}{{L + 1}}} \right)^{\frac{L}{4}}}\mathop {\det }\limits_{1 \le j,n  \le L/2} (\sin (j{k_n })).
\end{eqnarray}
Using the symplectic Vandermode determinant evaluation
\begin{eqnarray}
&&\mathop {\det }\limits_{1 \le n,j \le L/2} (x_n^j - x_n^{ - j}) = \prod\limits_{n = 1}^{L/2} {x_n^{ - L/2}(1 - x_n^2)} \nonumber\\
&&\,\,\,\,\,\,\,\,\,\,\,\,\,\,\,\,\,\,\,\,\,\, \times \prod\limits_{m = n + 1}^{L/2}  ({x_n} - {x_m})(1 - {x_n}{x_m}) , \nonumber
\end{eqnarray}
we obtain
\begin{eqnarray}
&&C_\alpha ^{(0)} ={i^{ - L/2}} {\left( {\frac{1}{{2(L + 1)}}} \right)^{\frac{L}{4}}}\left[ {\prod\limits_{n = 1}^{L/2} {{{\rm{e}}^{ - i(L/2){k_n}}}(1 - {{\rm{e}}^{2i{k_n}}})} } \right.\nonumber\\
&&\left. {\,\,\,\,\,\,\,\,\,\,\,\,\,\,\,\,\,\,\,\,\,\,\,\, \times \!\! \prod\limits_{m  = n + 1}^{L/2}  \!\! {({{\rm{e}}^{i{k_n}}} - {{\rm{e}}^{i{k_m }}})(1 - {{\rm{e}}^{i{k_n}}}{{\rm{e}}^{i{k_m }}})} } \right] .
\end{eqnarray}
The corresponding $|C_\alpha ^{(0)}{|^2} $ reads
\begin{eqnarray}
&&|C_\alpha ^{(0)}{|^2} = {2^{\frac{L}{2}(L - 1)}}{\left( {\frac{1}{{L + 1}}} \right)^{\frac{L}{2}}}\left[ {\prod\limits_{n = 1}^{L/2} {{{\sin }^2}{k_n}} } \right.\nonumber\\
&&\left. { \times \prod\limits_{m = n + 1}^{L/2} {{{\sin }^2}\left( {\frac{{{k_n} - {k_m}}}{2}} \right){{\sin }^2}\left( {\frac{{{k_n} + {k_m}}}{2}} \right)} } \right] .
\end{eqnarray}
From these overlaps, we obtain Eq.~(\ref{equationDWopen}).

\subsection{Periodic boundary conditions}
For periodic boundary conditions, the (squared) overlaps for the domain wall are now
\begin{eqnarray}
|C_\alpha ^{({\rm{0}})}{|^2} = {\left( {\frac{1}{L}} \right)^{\frac{L}{2}}}{\left| {\mathop {\det }\limits_{1 \le j,n \le L/2} ({e^{ - i{k_n}j}})} \right|^2}
\label{CalphaDW}
\end{eqnarray}
To evaluate this determinant, it is convenient to use the Vandermonde determinant formula. We obtain
\begin{eqnarray}
\mathop {\det }\limits_{1 \le j,n \le L/2} ({e^{i{k_n}j}}) = \prod\limits_{n = 1}^{L/2} {\prod\limits_{m = n + 1}^{L/2} \!\!\! {\left( {{e^{i{k_m}}} - {e^{i{k_n}}}} \right)} } \prod\limits_{n = 1}^{L{\rm{/2}}} {{e^{i{k_n}}}} .
\label{detvandermonde}
\end{eqnarray}
Using Eq.~(\ref{detvandermonde}), expression~(\ref{CalphaDW}) becomes
\begin{eqnarray}
|C_\alpha ^{({\rm{0}})}{|^2} = {\left( {\frac{1}{L}} \right)^{\frac{L}{2}}}{\left| {\prod\limits_{n = 1}^{L/2} {\prod\limits_{m = n + 1}^{L/2} {\left( {1 - {e^{ i({k_n } - {k_m})}}} \right)} } } \right|^2}
\end{eqnarray}
Further simplification leads to
\begin{equation}
|C_\alpha ^{(0)} |^2 = \frac{1}{L^{\frac{L}{2}}} 2^{\frac{(L-2)L}{4}} \prod_{n = 1}^{L/2} \prod_{m= n+1}^{L/2} \sin^2 
\left(  \frac{ k_n - k_m }{2 }  \right) .
\label{overlapsDWXX}
\end{equation}
The survival amplitude then reads
\begin{eqnarray}
&&A(t) = {\left( {\frac{1}{L}} \right)^{\frac{L}{2}}}{2^{(L - 2)L/4}}\sum\limits_{{{\{ {k_n}\} } }} {\left[ {\prod\limits_{n = 1}^{L/2} {{{\rm{e}}^{ - i\cos \left( {k_n} t\right)}}} } \right.} \nonumber\\
&&\left. {\,\,\,\,\,\,\,\,\,\,\,\,\,\, \times \prod\limits_{n = 1}^{L/2} {\prod\limits_{m = n + 1}^{L/2} {{{\sin }^2}} } \left( {\frac{{{k_n} - {k_m}}}{{2}}} \right)} \right] .
\end{eqnarray}
Writing $A(t)$ in terms of real quantities only we obtain
\begin{eqnarray}
&&F(t)=\frac{2^{(L-2)L/2}}{L^L} \left| \sum_{\{ k_n \}}  \left\{ 
\cos \left[ t \sum_{n=1}^{L/2} \cos \left( k_n\right) \right]  \right. \right.
\nonumber \\
&&\left. \left. \prod_{n=1}^{L/2} \prod_{m=n+1}^{L/2}  \sin^2 \left( \frac{ k_n - k_m}{2}  \right)
\right\} \right|^2,
\label{fidelityDWXX}
\end{eqnarray}
where $\{ k_n \}$ is the set of $\left( {\begin{array}{*{20}{c}}
L\\
{L/2}
\end{array}} \right)$ elements consisting of all combinations of momenta corresponding to the first $L$ odd (even) numbers in $\frac{L}{\pi } \times \{1, 3, \ldots, 2L-1\}$ (in $\frac{L}{\pi } \times \{2, 4, \ldots, 2L\}$) if $a=1$ ($a=0$). In the thermodynamic limit, the survival probability decay approaches the Gaussian behavior given by $F(t) = \exp (-t^2/2)$, which differs from the case of periodic boundary conditions by a factor of $1/2$ (see Sec.~\ref{DW}).


\end{document}